\documentclass[]{JHEP3}
\usepackage{epsfig,psfrag}

\newcommand{\simlt}{\stackrel{<}{{}_\sim}}
\newcommand{\simgt}{\stackrel{>}{{}_\sim}}
\newcommand{\gmsb}{$\tilde{g}$MSB }
\newcommand{\kahler}{K\"{a}hler }
\newcommand{\ci}{C^{5_1 5_2} }
\newcommand{\coj}{C^{5_1}_{j} }
\newcommand{\cof}{C^{5_1}_{3} }
\newcommand{\ctj}{C^{5_2}_{j} }
\newcommand{\cto}{C^{5_2}_{1} }
\newcommand{\ctt}{C^{5_2}_{2} }
\newcommand{\ctf}{C^{5_2}_{3} }

\newcommand{\Ro}{R_{5_1} }
\newcommand{\Rt}{R_{5_2} }

\newcommand{\go}{g_{5_1} }
\newcommand{\gt}{g_{5_2} }

\title{Phenomenology of Twisted Moduli in Type I String Inspired Models}

\author{B. C. Allanach\\ LAPTH, 9 Chemin de Bellevue, B.P. 110,
F-74941, Annecy-le-vieux, FRANCE.\\ E-mail: \email{allanach@cern.ch}}
\author{S. F. King\\ Department of Physics and Astronomy, University
of Southampton, Southampton, SO17 1BJ, UNITED KINGDOM.\\ E-mail:
\email{sfk@hep.phys.soton.ac.uk}}
\author{D. A. J. Rayner\\ Dipartimento di Fisica `G. Galilei',
Universit\'{a} di Padova and INFN, Sezione di Padova, Via Marzola 8,
I-35131 Padua, ITALY.\\ E-mail: \email{rayner@pd.infn.it}}

\abstract{We make a first study of the phenomenological
implications of twisted moduli in type I intersecting D5-brane
models, focussing on the resulting predictions at the LHC
using SOFTSUSY to estimate the Higgs and sparticle spectra.
Twisted moduli can play an important role in giving
a viable string realisation of sequestering in the limit
where supersymmetry breaking comes entirely from the 
twisted moduli.  We focus on
a particular string inspired version of gaugino mediation in 
which the first two families are localised at the intersection
between D5-branes, whereas the third family and Higgs 
doublets are allowed to move within the world-volume of one of the
branes.  The soft supersymmetry breaking third family sfermion mass 
terms are then in general non-degenerate with the first two families. 
We place constraints upon parameter space and predictions of flavour
changing neutral current effects.  Twisted moduli domination is studied  
and, as well as solving the most serious part of the SUSY flavour problem,
is shown to be highly constrained.  The constraints are 
weakened by switching on gravity-mediated contributions from the
dilaton and untwisted T-moduli sectors.  In the twisted moduli
domination limit we predict a stop-heavy MSSM spectrum and
quasi-degenerate lightest neutralino and chargino states with
wino-dominated mass eigenstates.}

\keywords{D-branes, Supersymmetry Breaking, Supersymmetry Phenomenology}

\preprint{SHEP 04/09\\ DFPD-04/TH/07\\ hep-ph/0403255}

\begin{document}

%%%%%%%%%%%%%%%%%%%%%%%%%%%%%%%%%%%%%%%%%%%%%%%%%%%%%%%%%%
%%%%%%%%%%%%%%%%%%%%%%%%%%%%%%%%%%%%%%%%%%%%%%%%%%%%%%%%%%
%%%%%%%%%%%%%%%%%%%% SECTION: INTRO %%%%%%%%%%%%%%%%%%%%%%
%%%%%%%%%%%%%%%%%%%%%%%%%%%%%%%%%%%%%%%%%%%%%%%%%%%%%%%%%%
\section{Introduction}  \label{sec:intro}

Superstrings provide a consistent way of unifying the
four fundamental forces of Nature together within a single
calculable theory.  Until recently the weakly-coupled heterotic
string theories were believed to offer the most promising framework in
which to recover the observable low-energy Standard
Model physics.  However there was a major paradigm shift following the
discovery of string dualities~\cite{dual} that identified the
separate string theories as different perturbative limits of an underlying
11d M-theory~\cite{mtheory}.  In particular the type I theory was
recognised as a viable alternative to heterotic models, but with novel features
such as a variable fundamental string scale $M_{\ast}$ and the
presence of solitonic Dirichlet branes (D-branes) in the low-energy
spectrum~\cite{dbranes}. 
The flexibility of the string scale allows for models with large
extra-dimensions~\cite{largexd} which open up the possibility of observing
``stringy'' experimental signatures at future colliders, and also
non-standard gauge coupling unification at scales well below $M_X \sim
2 \times 10^{16}$ GeV~\cite{xduni}.  The Dirichlet branes are required for
consistency, but provide a mechanism for localising matter and gauge fields
(open string states) on a lower-dimensional slice of the full 10d
spacetime in which gravity (closed strings) lives.  
These exciting new features have lead to a renaissance in string phenomenology,
and renewed attempts to derive the (supersymmetric) Standard Model
directly from a string theory compactification.

Recent developments have also motivated the study of
higher-dimensional field theories that may
describe the low-energy limit of string theory.  Many
string-inspired ideas and techniques including branes, orbifolding and
Kaluza-Klein towers are now commonly used in model-building~\cite{xdrev}. These
models are studied with effective field theory (EFT) techniques valid up to
some ultraviolet (UV) cutoff scale $\Lambda_{UV} < M_{\ast}$, where the
string-like features cannot be resolved.  In particular, branes can be
associated with orbifold fixed points rather than the string-theoretic
picture of hypersurfaces where open strings end, and strings are
treated as point-particles.  Also Standard Model
states can be allocated arbitrarily to different points/branes in the
higher-dimensional space, and often orbifold symmetry assignments
dictate whether a particular coupling is allowed or not~\cite{xdrev2}.  

Many new models have been constructed to investigate quasi-realistic extensions
of the (supersymmetric) Standard Model, where the extra-dimensional
framework offers novel solutions to problems in cosmology, grand
unification and flavour physics.  In particular supersymmetry (SUSY)
breaking has received a lot of attention in the context of higher-dimensional
parallel-brane models.  In these models a pair of parallel
4d branes are localised at either end of a 5d (or higher) bulk
spacetime as shown in Figure \ref{fig:gmsb}.  The
Standard Model fields are localised on one brane while SUSY is broken
at a distance on the other.  These models are inspired by the
strongly-coupled heterotic limit of 11d M-theory~\cite{hw}, and offer an
attractive realisation of hidden-sector SUSY breaking.
In the original hidden-sector models, the visible and SUSY breaking sectors
occupied the 
same 4d spacetime, but direct communication between sectors was suppressed
by inverse Planck-scale couplings.  In contrast, the new parallel-brane
models avoid direct couplings by delocalising -- or ``sequestering'' --
the two sectors at different points within the higher-dimensional
space.  Thus communication of SUSY breaking requires
additional intermediary fields living in the bulk that couple to
both sectors, and anomaly (gravity)~\cite{amsb} and gaugino~\cite{gmsb}
mediation are two recent examples.  The combined \kahler
potential that describes the visible, hidden and bulk sectors is
assumed to have a particular sequestered form which is reminiscent of
no-scale gravity models~\cite{noscale}. 

SUSY breaking masses and trilinears for Standard Model fields on the
visible brane arise from higher-dimensional loops of particles living
in the 5d bulk, since the contributions from direct couplings between
sectors are found to be exponentially-suppressed by the separation $r$
and cutoff scale $\Lambda_{UV}$.  The absence of significant direct couplings
leads to an automatic suppression of flavour changing neutral currents
(FCNCs) which alleviates the SUSY flavour problem since small off-diagonal
squark/slepton mass-matrix elements are only reintroduced at the weak-scale
through renormalisation group equation (RGE) running effects down from
$\Lambda_{UV}$.  However in the simplest models the viable regions of
parameter space are very constrained.  Also the special form of \kahler
potential that leads to sequestering has been criticised as unlikely in
explicit string constructions~\cite{dine} and therefore the
sequestered parallel-brane models are believed to be unrealistic.

Our goal in this work is to tackle these problems by embedding
sequestering in a type I string model of the Minimal Supersymmetric Standard
Model (MSSM) involving intersecting Dirichlet branes that may avoid
the problems identified in Ref.~\cite{dine}.  We realise the
sequestering mechanism by identifying the hidden sector with localised
closed string twisted moduli states $Y_k$~\cite{benakli,tmsb} that
acquire SUSY breaking F-term vacuum expectation values (VEVs).  These
twisted moduli states can be 
spatially-separated away from (visible sector) MSSM fields that are
trapped at the intersection between different D-branes with apparently no
direct coupling to the twisted moduli.  There are additional gravity
mediation contributions to SUSY breaking from delocalised closed
string states (dilaton and untwisted T-moduli) that freely move in the
full 10d spacetime.  In order to model the sequestering effect, we have
previously proposed a modified \kahler potential for open string
states that stretch between two different D5-branes and
are localised away from the SUSY breaking twisted moduli
fields~\cite{tmsb}.  
% \cite{ibanez94,ibanez97,ibanez98}
In the absence of a complete theory of SUSY breaking we parameterise the
F-terms using Goldstino angles~\cite{ibanez94}-\cite{ibanez98}
which control the relative contributions to the overall SUSY breaking,
and we derive explicit expressions for the soft Lagrangian
parameters~\cite{susyrev}.

We will consider an intersecting D5-brane model of the R-parity
conserving MSSM~\cite{shiutye,bmsb} where the first two MSSM families
are trapped at the 
intersection between D-branes and the third family can couple directly
to the twisted moduli.  We consider the ``single-brane dominance''
limit with the MSSM gauge groups dominated by their components on the
D$5_2$-brane, and use SOFTSUSY~\cite{softsusy} to perform the
RGE analysis below the grand unification scale of
$M_X \sim 2 \times 10^{16}$ GeV.  We will focus our study around the
twisted moduli domination limit where the only source of
SUSY breaking originates from the localised twisted moduli fields, and
this limit corresponds to sequestered models like gaugino mediated
SUSY breaking (\gmsb\hspace*{-2mm})~\cite{gmsb}.  However we will also
study the effect of  
including gravity mediation by smoothly varying the Goldstino angles,
and we observe that the highly-constrained sequestered models are much
less constrained after the inclusion of additional gravity mediated effects.
The most severe FCNC 
constraints involving the first two families are satisfied within our
model in two different ways.  In the limit of twisted moduli
domination the first and second family soft sfermion masses are
negligible at the high-scale, and are only generated at low-energies by
flavour-blind RGE effects.  The inclusion of gravity mediation spoils
this mechanism for suppressing FCNCs since soft sfermion masses are no
longer negligible at the high-scale.  Instead the SUSY flavour problem is
ameliorated with the combination of heavy gluinos and the assumption
of family-diagonal soft scalar mass matrices.

There have been many previous phenomenological analyses of type I
D-brane constructions that use two Goldstino angles $\theta$, $\phi$
to parameterise the relative contributions to SUSY breaking from the
dilaton, twisted and untwisted moduli F-term VEVs.  However we present the
first analysis to focus on the twisted moduli domination limit by
proposing a \kahler potential to model the sequestering mechanism.
Unlike previous studies with only D9-branes~\cite{ben,ben2}, we 
consider a model involving intersecting D5-branes that enable open
string MSSM states to be trapped at sub-manifolds of the full 10d
spacetime that can be spatially-separated away from the localised
twisted moduli source of SUSY breaking.  Although twisted moduli arise
in D9-brane constructions, it is impossible to sequester open
string states away from the SUSY breaking since the open
strings are free to move throughout the full D9-brane world-volume.
Therefore we would not expect any sequestering to occur as these open strings
can couple directly to the twisted moduli, and the standard \kahler
potentials of Ref.~\cite{ibanez98} are valid.  

The layout of the remainder of the paper is as follows.  In section
\ref{sec:sequestering} we review our string-inspired gaugino mediation
model of sequestering with twisted moduli.  We also
list the SUSY breaking soft parameters and motivate our choice of
model parameters.  Our results are presented in section
\ref{sec:results} for the twisted moduli domination limit, and also
more general cases involving the dilaton and untwisted T-moduli
contributions.  We also discuss constraints on our model from flavour
changing processes and highlight some ``benchmark'' points with
sample spectra.  Our conclusions and discussion are given in section
\ref{sec:disc}.  For 
completeness we derive the SUSY breaking F-terms using Goldstino
angles in Appendix \ref{app:softterms}, and discuss the
phenomenological problems facing a similar model with three degenerate
families of intersection states in Appendix \ref{app:gmsb}.

%%%%%%%%%%%%%%%%%%%%%%%%%%%%%%%%%%%%%%%%%%%%%%%%%%%%%%%%%%
%%%%%%%%%%%%%%%%%%%%%%%%%%%%%%%%%%%%%%%%%%%%%%%%%%%%%%%%%%
%%%%%%%%%%%%%%% SECTION: OUR MODEL (REVIEW) %%%%%%%%%%%%%%
%%%%%%%%%%%%%%%%%%%%%%%%%%%%%%%%%%%%%%%%%%%%%%%%%%%%%%%%%%
%\newpage
\section{Sequestering with twisted moduli}  \label{sec:sequestering}
 
In this section we will review the results from our earlier
paper~\cite{tmsb} where we considered the contribution to SUSY
breaking soft parameters from the F-term VEV of localised twisted
moduli fields $Y_k$.  Motivated by \gmsb and non-perturbative
instanton effects we proposed a modified \kahler potential for
intersection states $\ci$ that offers a string realisation of
sequestering.  Our framework allows us to study sequestering
in the presence of other F-term VEVs from the dilaton and untwisted
moduli fields - in this way we can study the interplay between \gmsb
($F_{Y_k} \neq 0$) and pure gravity mediation ($F_S , F_{T_i} \neq 0$)
contributions.  We will see that the strong constraints applying to
sequestered models are relaxed through the addition of gravity
mediation effects. 

Our starting point is the low-energy effective supergravity (SUGRA)
description of 
type I models compactified on toroidal orbifolds (or type IIB
orientifolds)~\cite{ibanez98} which involve two stacks of
perpendicular intersecting 
D5-branes at the origin fixed point as shown in Figure \ref{fig:gmsb}.  
%\vskip-5mm
\FIGURE[h]{
 \epsfig{file=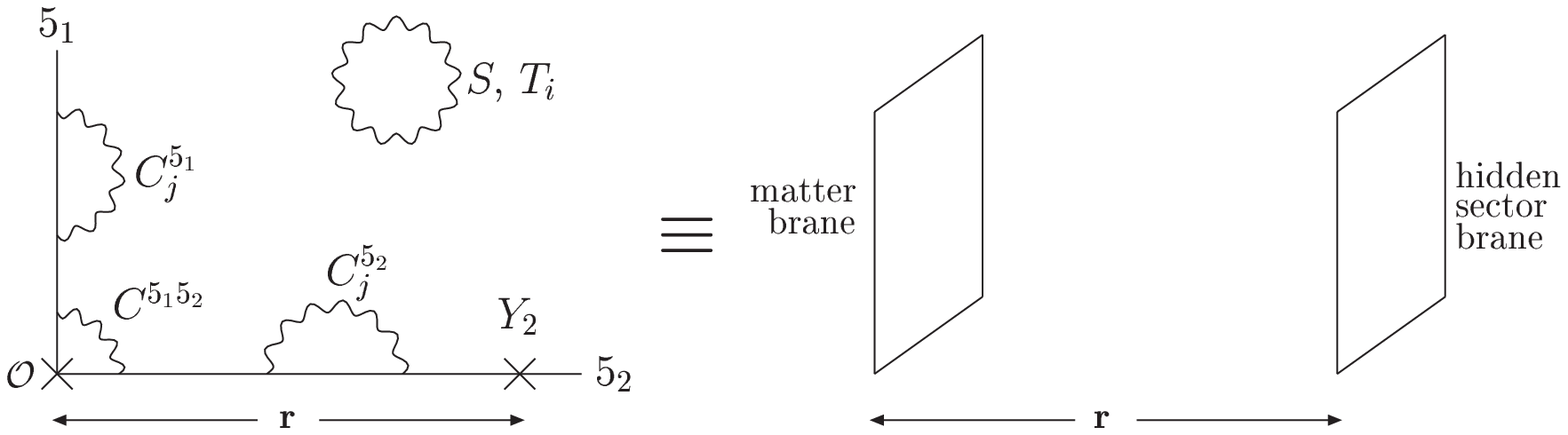,height=4.2cm}
 \caption{({\it left}) A generic type I construction involving
(perpendicular) intersecting D5-branes that share a common 4D overlap
at the origin fixed point $\cal O$, where each D-brane carries a corresponding
gauge group.  The open and closed string states are shown, including
twisted moduli that are trapped at orbifold fixed points marked by crosses.  We
propose that higher-dimensional field theory models ({\it right}) involving
branes/fixed points are analogous to our string theory setup, where the
matter (hidden sector) branes are equivalent to open strings at
the origin $\ci$ (twisted moduli) respectively.}
 \label{fig:gmsb}
}
Each stack of $N$ coincident D-branes supports a $U(N)$ super
Yang-Mills gauge group where gauge fields
arise as open strings with both ends attached to the same stack and
carrying adjoint quantum numbers~\footnote{For simplicity we will
assume that the 
generalised Green-Schwarz mechanism~\cite{gs} cancels any problematic
anomalous $U(1)$ factors.}.  Chiral matter fields are 
open strings carrying (bi)fundamental charges, with either (i) both
ends attached to the same brane~\footnote{The extra world-sheet
fermion index $j$ determines allowed couplings (in the renormalisable
superpotential $W_{ren}$) that arise from the splitting and joining of
different open string states subject to string-selection rules.},
e.g. $\coj$ and $\ctj$; or alternatively (ii) open strings stretched between
different D5-branes, e.g. $\ci$ states.  The string tension
effectively localises the intersection states $\ci$ at the
common 4d overlap region $\cal O$, which is in contrast to the other
open string states $\coj$ and $\ctj$ that can freely move within the full
6d world-volume of their corresponding D5-brane.  In addition to the
open strings, there are also closed string states including
three untwisted T-moduli $T_i$ (that parametrise the size of the
extra-dimensional geometry since $(T_i + \overline{T}_i) \sim
R_{5_i}^2$) and the dilaton $S$ that are not confined to the
world-volume of D-branes, but 
live in the full 10d space~\footnote{Notice that for models involving D9-branes
open string states also live in 10d, but these states are still
distinct from closed strings since their ends remain attached to the
D-branes.}. 
However closed strings can also be
localised in lower-dimensional sub-manifolds through the action of
orbifolding, and twisted moduli states $Y_k$ are closed strings
trapped around orbifold fixed points.  In this work we will assume that SUSY
breaking originates only from the closed string sector, i.e. when the
auxiliary F-terms of the chiral superfields containing the moduli and
dilaton acquire non-zero VEVs, i.e. $F_S$, $F_{T_i}$ and/or $F_{Y_k}
\neq 0$.  
%\vskip-5mm
\FIGURE[h]{
 \epsfig{file=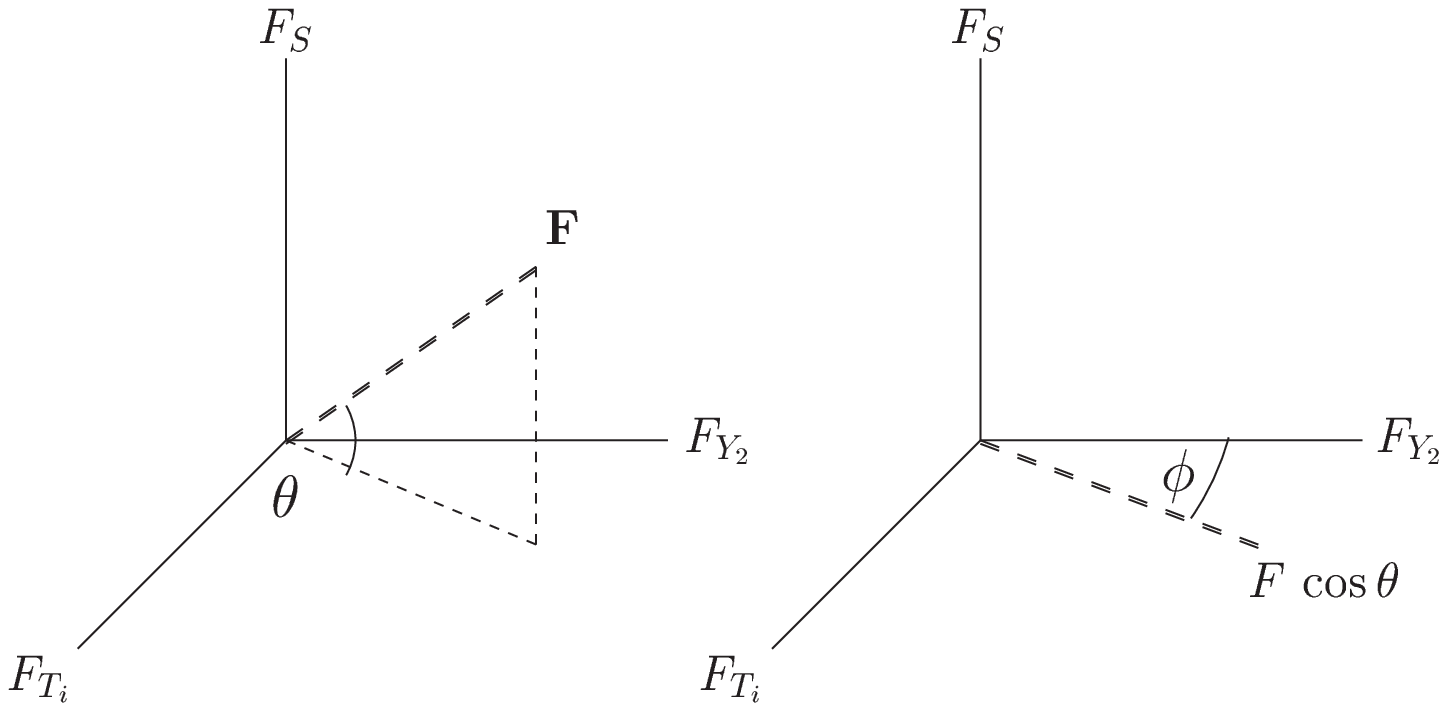,height=5cm}
 \caption{The relative contributions to the total SUSY
    breaking F-term ${\mathbf F}$ from each closed string F-term VEV
    ($F_S \, , \, F_{T_i} \, , \, F_{Y_2}$) can be parameterised by
    two Goldstino angles $(\theta \, , \, \phi)$
% \cite{ibanez94,ibanez97,ibanez98}
    \cite{ibanez94}-\cite{ibanez98}.} 
 \label{fig:fterms}
}

In the absence of a complete theory of SUSY breaking, such as gaugino
condensation~\cite{abel,kobayashi}, we represent the
different F-term VEVs as components of an overall F-term vector
${\mathbf F}$ as shown in Figure \ref{fig:fterms}.   Varying the two
Goldstino angles ($\theta, \,\phi$) 
allows us to control the relative contributions to SUSY breaking from
each closed string sector~\footnote{Although we will introduce
modular-anomaly cancelling Green-Schwarz terms~\cite{gs} that mix the
angular dependence of the twisted and untwisted F-terms from
Eq. \ref{eq:ftermdef}, we will still refer to the untwisted T-moduli
domination limit as $\cos\theta = \sin\phi=1$, and similarly for the
twisted moduli domination limit.}.
\begin{eqnarray}
 {\mathbf F} = \left(
  \begin{array}{c}
   F_S \\
   F_{T_i} \\
   F_{Y_k}
  \end{array} \right)
 \sim \left(
  \begin{array}{c}
   F \, \sin\theta \\
   F \, \cos\theta \, \sin\phi \, \Theta_i \\
   F \, \cos\theta \, \cos\phi \, \chi_k 
  \end{array} \right)
    \label{eq:ftermdef}
\end{eqnarray}

\noindent where $F$ is the absolute magnitude of the total F-term
${\mathbf F}$.  In this work we will only consider a single twisted
moduli $Y_2$ and set $\chi_2=1$, $\chi_{k \neq 2} = 0$ and we choose
$\Theta_i=1/\sqrt{3}$ for simplicity.  

The D-brane construction in the left panel of Figure \ref{fig:gmsb}
shows that it is possible to localise matter fields (open strings
$\ci$) and SUSY breaking fields (twisted moduli $Y_2$) at {\bf different} fixed
points in the compactified space~\footnote{There may also be twisted
moduli fields trapped at the same origin fixed point as the $\ci$
states, but we will assume that they play no r\^{o}le in SUSY breaking in
what follows.}.  In the limit that the only non-zero F-term VEV arises from
the localised twisted moduli $Y_2$, the setup was
observed~\cite{benakli, bmsb}
to be similar to the recent higher-dimensional field theory orbifold
models of sequestering (right panel of Figure
\ref{fig:gmsb}).  In these orbifold models, matter fields and SUSY
breaking can be arbitrarily localised at different fixed
points/branes with apparently no direct coupling between the different
sectors~\cite{amsb,gmsb}.  Supersymmetry breaking can be communicated to
the sequestered matter fields through the super-conformal
anomaly~\cite{amsb} or transmission of bulk gauginos~\cite{gmsb} at
the 1-loop level, whereas the non-renormalisable higher-dimensional
operators that give direct couplings are exponentially-suppressed after
integrating out physics above the cutoff scale $\Lambda_{UV}$. 

Following the ideas of Refs~\cite{benakli,bmsb}, we propose that twisted
moduli offer a mechanism for realising sequestering in type I D-brane
constructions~\footnote{Recent work has disputed whether the 
sequestered \kahler potential is viable in string theory~\cite{dine},
but these authors did not consider exploiting localised twisted moduli
fields.} between localised SUSY breaking fields ($Y_2$) and spatially-separated
matter fields ($\ci$ and $\coj$).  However our D-brane framework also
allows for some open strings states ($\ctj$) to couple directly to the
twisted moduli F-term VEV, and this can be exploited to generate a
hierarchy between the soft scalar masses of different states at the
high-scale.  In the next section we will 
outline a string-inspired MSSM model where the third
family and Higgs doublets are identified with $\ctj$ states and 
consequently have hierarchically larger soft masses (at the
high-scale) compared to the first two MSSM generations which are
sequestered at the origin as $\ci$ states.  We will study the
phenomenology of this model as we move away smoothly from the twisted
moduli dominated limit ($\cos\theta=\cos\phi=1$) and introduce
contributions to the SUSY breaking from the dilaton and T-moduli
F-terms.  Unlike previous analyses~\cite{ben,ben2}, we will focus on
the twisted moduli domination limit.  We will show that the strong
constraints applying to that limit are relaxed through the addition of
gravity mediation effects. 

%%%%%%%%%%%% string-inspired model %%%%%%%%%%%%%%%%%%%%
\subsection{String-Inspired Gaugino Mediation (SI$\tilde{g}$M)}  
  \label{sec:sigma}

In this section we will outline our model of sequestering as shown in Figure
\ref{fig:modelb}.  We choose to study this
particular setup because it is the closest string-inspired model to
gaugino mediation in the literature.  It is important to emphasise
that a model where all three MSSM
families, rather than just the first two, are intersection states is
not directly motivated by string constructions, moreover if such a
model is enforced by hand it leads to phenomenological problems as
discussed in Appendix \ref{app:gmsb}.  
%\vskip-7mm
\FIGURE[h]{
 \epsfig{file=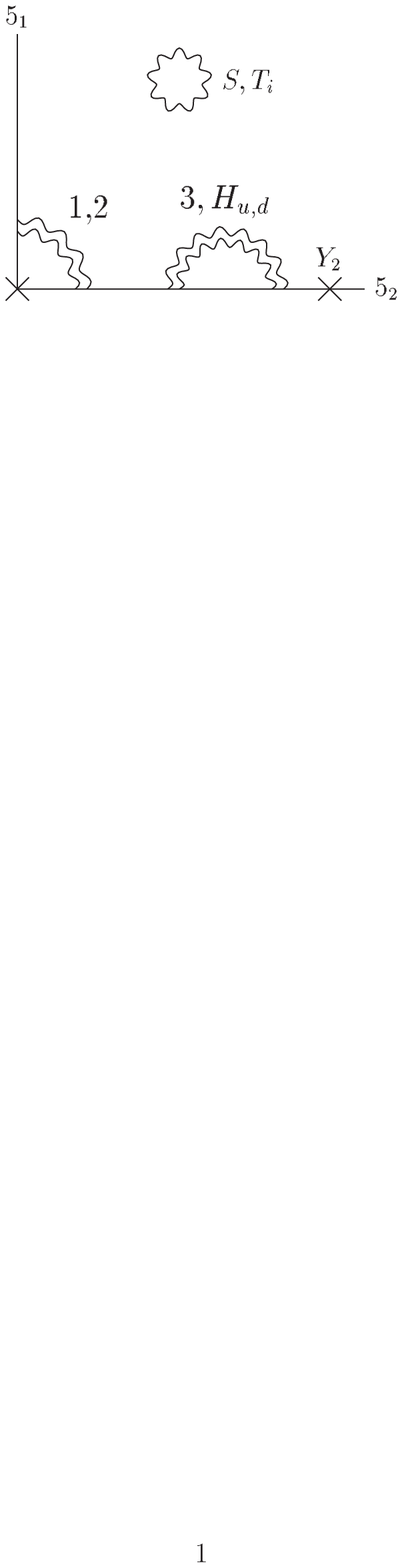,height=4.5cm}
 \caption{The allocation of open and closed string states in
a string-inspired MSSM model.  The first/second family ($1,2$) are
intersection states $\ci$, while the third family ($3$) and Higgs
doublets ($H_u, H_d$) are $\ctj$ states.  The twisted moduli $Y_2$ is
trapped at a fixed-point within the D$5_2$-brane, whereas the dilaton
$S$ and untwisted moduli $T_i$ live in the full 10d spacetime.}   
 \label{fig:modelb}
}

The model studied here suppresses the most dangerous FCNCs involving
the first and second families using two different mechanisms.  Firstly
in the limit of twisted moduli domination, the first/second family
scalar masses are negligible at the high-scale due to sequestering.
This leads to universal low-energy squark and slepton mass matrices
that are approximately diagonal due to 
flavour-blind RGE running effects in analogy to \gmsb models.  Moving
away from the twisted moduli limit by introducing gravity mediation
contributions removes the sequestering, and so the first/second family soft
scalar masses are no longer negligible at the high scale which spoils
this mechanism for suppressing FCNCs.  However, we
find that the physical gluino and squark sparticle masses are
sufficiently heavy to ameliorate the FCNC constraints when we make the
additional assumption that the soft scalar mass matrices are diagonal in
family-space at the high-scale.  The weaker experimental constraints
on the third family FCNCs allows us to treat these fields
differently.  Direct coupling to the twisted moduli gives
heavier third family soft masses at the high-scale that prove helpful
in achieving successful electroweak symmetry breaking (EWSB) and a sufficiently
heavy lightest CP-even Higgs mass.

Our setup involves two stacks of
perpendicular intersecting D5-branes where a copy of the MSSM gauge
group $G_{MSSM}$ lives on both stacks.  We assume a hierarchy
between the different toroidal compactification radii $\Rt \gg \Ro$
which implies that the corresponding gauge couplings on each brane are also
hierarchical with $\gt \ll \go$.  This assumption allows us to treat
the $5_1$-brane (and any attached open string states) as effectively
localised at the 4d intersection region between branes.  Also the
resulting hierarchy between gauge couplings ensures that the diagonal
product of MSSM gauge groups is dominated by the components living
within the $5_2$-brane. This ``single brane dominance'' limit allows for
gauge coupling unification on the $5_2$-brane.
However the $5_1$-brane could not be ignored from the start since it
facilitates the localisation of open string states $\ci$ at the 4d
intersection which we identify with the first and second
quark/lepton families.  In contrast the third
family and Higgs doublets are allocated to open string states $\ctj$ with
both ends attached to the $5_2$-brane~\footnote{This
construction is a simplification of an existing $Z_6$-orbifold
compactification in the presence of a background
flux~\cite{shiutye,bmsb}, where the Pati-Salam group arises on both
stacks of branes, and three families are divided between the $\ci$ and
$\ctj$ open string states.}.  We choose the following
allocation of MSSM superfields:
\begin{eqnarray}
 Q_{iL} \, , \, L_{iL} \, , \, U_{iR}^c \, , \, D_{iR}^c 
  \, , \, E_{iR}^c &\equiv& C^{5_1 5_2} \hspace*{1cm} (i=1,2) 
    \label{eq:openall}  \\
 U_{3R}^c \, , \, D_{3R}^c \, , \, E_{3R}^c \equiv C^{5_2}_3
  \hspace*{5mm} , \hspace*{5mm} 
 H_{u} \, , \, H_d &\equiv& C^{5_2}_2
  \hspace*{5mm} , \hspace*{5mm}
 Q_{3L} \, , \, L_{3L} \equiv C^{5_2}_1  \nonumber
\end{eqnarray}
At the renormalisable level, the superpotential is constrained by
string selection rules~\cite{ibanez98} to have the following allowed
terms (subject to gauge-invariance)~\footnote{The coefficients in
Eq.(\ref{eq:wren}) are of order $g_{5_2}$ leading to approximate
Yukawa unification for the third family.  There will be further
corrections coming from higher-order operators in a more complete
theory of flavour which may relax Yukawa unification further.}:
\begin{eqnarray}
 W_{ren} \sim \cto \, \ctt \, \ctf + \ctf \, \ci \, \ci   \label{eq:wren}
\end{eqnarray}
and the allocation of states in Eq.(\ref{eq:openall}) leads to
hierarchical Yukawa (and trilinear) matrices with a non-zero (33)
entry, and the other (smaller) elements of the Yukawa matrices
required to generate the other fermion masses are assumed to arise
from higher-dimensional operators.  We will also impose an R-parity (or
gauge-invariance if the model is embedded within a larger gauge group)
to forbid the R-parity violating string-allowed superpotential terms
given by the second term in Eq.(\ref{eq:wren}).  Notice that the 
renormalisable superpotential contains no Higgsino mixing term $\mu H_u H_d$.

In a previous paper~\cite{tmsb} we discovered that a straightforward
application of 
low-energy supergravity formulae for soft parameters, using the
standard type I \kahler potentials for open string
states~\cite{ibanez98} and Goldstino parameterisation of SUSY breaking,
does not lead to sequestered first/second 
family ($\ci$) soft scalar masses in the limit of twisted moduli
domination.  In fact $m_{\ci}^2$ is independent of the separation 
$r \sim {\cal O}(\Rt)$ between the SUSY breaking $Y_2$ and the D5-brane
intersection.  In Ref.~\cite{tmsb} we proposed a modified
\kahler potential $K_{\ci}$ for $\ci$ states with an explicit
dependence on $r$ that, by construction, exhibits
sequestering in the twisted moduli domination limit.  Note that we have
chosen a sequestering exponential factor 
$e^{-(M_{\ast} R_{5_2})^2} = e^{-(T_2+\overline{T_2})/4}$ which is
different from the usual Yukawa-like factor in the parallel-brane
sequestered models, since we assume that this suppression factor
originates from a non-perturbative instanton effect due to strings
stretching between different fixed points~\cite{altexp}.  The relevant
terms in the \kahler potential for the open string states in
Eq.(\ref{eq:openall}) are~\footnote{Throughout we will suppress powers
of the fundamental string scale $M_{\ast}$ with the understanding that
the VEVs of $S, T_i$ and $Y_2$ are in units of $M_{\ast}$.}:
\begin{eqnarray}
 K_{C^{5_2}_{1}} = \frac{|\cto|^2}{T_{3}+\overline{T_3}} 
  \hspace*{1cm}, \hspace*{1cm}  
 K_{C^{5_2}_{2}} = \frac{\, |\ctt|^2 \,}{S+\overline{S}} 
  \hspace*{1cm}, \hspace*{1cm}  
 K_{C^{5_2}_{3}} = \frac{|\ctf|^2}{T_{1}+\overline{T_1}}
    \label{eq:matk} \\
 K_{C^{5_1 5_2}} =
  \frac{|\ci|^2}{\sqrt{S+\overline{S}} \, \sqrt{T_{3}+\overline{T_{3}}}} \,
   exp \left[ \frac{X^2}{6} \left( 1-e^{- (T_2 + \overline{T_2})/4} \right) 
    \right]  \label{eq:matk2}  \hspace*{15mm}
\end{eqnarray}
where we have assumed that all K\"{a}hler potentials are
``canonical'', i.e. diagonal~\footnote{We refer the reader to
Ref.~\cite{peddie} for a recent discussion of canonical \kahler
potentials.}; and the modified $\ci$ potential 
involves the combination $X = Y_2 + \overline{Y_2}  
 - \delta_{GS} \, {\mathrm ln} ( T_2 + \overline{T_2} )$ due to
modular anomaly cancellation arguments~\cite{gs}.  In the limit of
vanishing compactification radius
$\Rt = \sqrt{ T_2 + \overline{T_2} }/2 \longrightarrow 0$, we
recover the usual expression for $K_{\ci}$ ~\cite{ibanez98}.  For the
closed string dilaton and moduli states, we use the standard
K\"{a}hler potentials:  
\begin{eqnarray}
 K(S, \overline{S}, T_i, \overline{T_i}, Y_2, \overline{Y_2})
  = - \ln (S + \overline{S}) - \sum_{i=1}^{3} \ln (T_i + \overline{T_i}) 
   + \hat{K} \left[ Y_2 + \overline{Y_2} 
    - \delta_{GS} \, \ln ( T_2 + \overline{T_2} ) \right] 
     \label{eq:kp} \hspace*{5mm}
\end{eqnarray}  
but we choose to leave the precise form of the twisted moduli K\"{a}hler
potential $\hat{K}$ as an unknown function with an argument
$X = Y_2 + \overline{Y_2} - \delta_{GS} \, {\mathrm ln} ( T_2 +
\overline{T_2} )$ ~\cite{twistkp}.  We will choose to fix the
first and second derivatives $\hat{K}' \equiv \partial_X \hat{K}(X) =
0$ and $\hat{K}'' \equiv \partial_{X}^{2} \hat{K}(X) = 1$ as discussed
in section \ref{sec:choose}.  However the
mixing induced between $T_2$ and $Y_2$ by $\hat{K}(X)$
means that the K\"{a}hler metric is no longer diagonal (after ignoring 
negligible VEVs for the matter fields).  Therefore, in order to invert
the K\"{a}hler metric, we use the same techniques as
Ref~\cite{ben,ben2} to define the SUSY breaking with Goldstino angles
where $F_{T_2}$ and $F_{Y_2}$ are expanded in inverse powers of the
$T_{2}+\overline{T_2}$ VEV as discussed in Appendix \ref{app:softterms}.

%%%%%%%%%%%%%%%%%%%%%%%%%%%%%%%%%%%%%%%%%%%%%%%%%%%%%%%%%
%%%%%%%%%%%%%%%%%%%%%%%%%%%%%%%%%%%%%%%%%%%%%%%%%%%%%%%%%
\subsubsection{SUSY breaking parameters}  \label{sec:sbp}

The soft Lagrangian ${\cal L}_{soft}$ is given by:
\begin{eqnarray}
 - {\cal L}_{soft}^{MSSM} = \frac{1}{2} \left( M_{1} \tilde{B} \tilde{B}
  + M_{2} \tilde{W} \tilde{W} + M_{3} \tilde{g} \tilde{g} \right)
   + h.c. \hspace*{6cm} \nonumber \\
 + (A^{u}_{ij} \, Y^{u}_{ij}) \, \tilde{u}^{\ast}_{iR} \tilde{Q}_{jL} H_{u} 
  - (A^{d}_{ij} \, Y^{d}_{ij}) \, \tilde{d}^{\ast}_{iR} \tilde{Q}_{jL} H_{d} 
   - (A^{e}_{ij} \, Y^{e}_{ij}) \, \tilde{e}^{\ast}_{iR} \tilde{L}_{jL} H_{d}
    + h.c. \hspace*{1.5cm} \label{eq:lsoft} \\
 + \tilde{Q}_{iL}^{\dagger} \left(m_{\tilde{Q}}^{2}\right)_{ij} \tilde{Q}_{jL}
 + \tilde{L}_{iL}^{\dagger} \left(m_{\tilde{L}}^{2}\right)_{ij} \tilde{L}_{jL}
 + \tilde{u}_{iR}^{\ast} \left(m_{\tilde{u}}^{2}\right)_{ij} \tilde{u}_{jR}
 + \tilde{d}_{iR}^{\ast} \left(m_{\tilde{d}}^{2}\right)_{ij} \tilde{d}_{jR}
  \hspace*{0mm} \nonumber \\
 + \tilde{e}_{iR}^{\ast} \left(m_{\tilde{e}}^{2}\right)_{ij} \tilde{e}_{jR}
  + m_{H_{u}}^{2} H_{u}^{\ast} H_{u} + m_{H_{d}}^{2} H_{d}^{\ast} H_{d}
  - \left( B \mu H_{u} H_{d} + h.c. \right)  \hspace*{-5mm} \nonumber
\end{eqnarray}
where ($\tilde{B}, \tilde{W}, \tilde{g}$) are the bino, wino and gluino
gauginos respectively; ($H_u, H_d$) are the scalar components of
the two Higgs doublets; and $\tilde{u}_{iR}, \tilde{d}_{iR},
\tilde{e}_{iR} \left( \tilde{Q}_{iL}, \tilde{L}_{iL} \right)$ are the
squark/slepton singlets (doublets) respectively.

We apply standard SUGRA formulae to Eqs.(\ref{eq:matk}-\ref{eq:kp})
- using the SUSY breaking F-terms from Eq.(\ref{eq:fterms3}) with
$k=1$ - to obtain expressions for the squark/slepton mass-squared matrices
$m^{2}$, the gaugino masses $M_{i}$, and the soft trilinear matrices
$A_{ij}$~\cite{susyrev}.  These soft parameters provide grand unified
theory (GUT) scale boundary 
conditions for the RGE analysis.  We do not
specify $B$ or $\mu$ at the GUT-scale since they can be exchanged for
$\tan\beta$ and the measured value of $M_{Z^0}$ by imposing radiative EWSB 
at the weak-scale. 

\vskip5mm
%\newpage
\noindent
$\bullet$ {\bf Soft scalar masses:}
\vskip2mm

Using the canonically-normalised \kahler potentials in
Eq.(\ref{eq:matk},\ref{eq:matk2}), we find that the squark and slepton
mass-squared matrices take the following form in family space at the
high-scale: 
\begin{eqnarray}
 m_{\tilde{Q}}^{2} \, , \, m_{\tilde{L}}^{2} \, , \, m_{\tilde{u}}^{2}
  \, , \, m_{\tilde{d}}^{2} \, , \, m_{\tilde{e}}^{2} = \left(  
   \begin{array}{ccc} 
    m_{\ci}^2 & m_{\ci}^2 & 0 \\
    m_{\ci}^2 & m_{\ci}^2 & 0 \\
    0 & 0 & m_{C^{5_2}_{1,3}}^2 
   \end{array}  \right)  \label{eq:ssms}
\end{eqnarray}
where the degenerate first/second family mass-squared is given by:
\begin{eqnarray}
 m_{C^{5_1 5_2}}^2 = m_{3/2}^2 \left[
  1 - \frac{3}{2} \sin^{2} \theta  
   - \frac{1}{2} \cos^{2} \theta \, \sin^{2} \phi  
 - \left( 1-e^{-(T_{2}+\bar{T}_{2})/4} \right) \cos^{2}\theta
  \cos^{2}\phi  \right. \label{eq:softci} \hspace*{15mm}\\
 - \frac{X}{3} \, \cos^2 \theta \sin^2 \phi \, \delta_{GS}
  \left( 1- e^{-(T_{2}+\bar{T}_{2})/4} \right) 
   + \frac{X^2}{96} \cos^2 \theta \sin^2 \phi \, 
    e^{-(T_{2}+\bar{T}_{2})/4} (T_{2}+\bar{T}_{2})^{2} \hspace*{5mm}
     \nonumber \\
 - \left. \frac{1}{16 \sqrt{3}} \cos^{2}\theta \cos\phi \sin\phi \, 
  e^{-(T_{2}+\bar{T}_{2})/4} \left\{ 8 (T_{2}+\bar{T}_{2}) +
  \delta_{GS} \, X \right\} X  \right] 
   + {\cal O} \left[ \frac{\delta_{GS} \, 
    e^{-(T_2+\overline{T_2})/4}}{(T_2+\overline{T_2})} \right]
\nonumber \hspace*{-5mm} 
\end{eqnarray} 
with $X = Y_2 + \overline{Y_2} 
 - \delta_{GS} \, {\mathrm ln} ( T_2 + \overline{T_2} )$; and the
third family mass-squared is
\begin{eqnarray}
 m_{C^{5_2}_1}^2 \, , \, m_{C^{5_2}_3}^2 
  = m_{3/2}^2 \left( 1 - \cos^2 \theta \, \sin^2 \phi \right)
    \label{eq:softc2j}
\end{eqnarray}
In the twisted moduli dominated limit, $m_{\ci}^2 \longrightarrow 0$
due to the exponential suppression.  Only diagonal mass matrix entries
are generated by gauge loops in the renormalisation to low energies.
Departing from twisted moduli domination introduces a non-negligible
$m_{\ci}^2$, and then we require an additional family symmetry to
suppress the $(1,2)$ entries at the GUT scale in order to avoid
predicting large FCNCs.  In practice then, for all numerical
estimates, we will use family diagonal boundary conditions for
sfermion soft masses. 

The Higgs doublet soft scalar masses-squared are universal at the
high-scale:
\begin{eqnarray}
 m_{H_u}^2 \, , \, m_{H_d}^2 \equiv
  m_{C^{5_2}_2}^2 = m_{3/2}^2 \left( 1 - 3 \sin^2 \theta \right) 
  \label{eq:softc22}
\end{eqnarray}
  
\noindent
$\bullet$ {\bf Soft gaugino masses:}
\vskip2mm

The soft gaugino masses explicitly depend on the gauge couplings.  In
this work we assume that all three Standard Model couplings 
unify to a common value at the usual GUT-scale $M_{GUT} \sim 2 \times
10^{16}$ GeV, and the unified gauge coupling is determined
by running the RGEs up from the weak-scale.  The soft masses are given by: 
\begin{eqnarray}
 M_{\alpha} = \frac{\sqrt{3} m_{3/2} \, g_{\alpha}^{2}}{8\pi} \cos \theta
  \left[ \frac{\sin \phi}{\sqrt{3}} 
   \left\{ T_{2} + \bar{T_{2}}
 + \frac{s_{\alpha}}{4\pi} \delta_{GS} \right\}
   \right. \label{eq:softga} \hspace*{5cm} \\
 - \left. \cos \phi \left\{ \frac{\delta_{GS}}{T_{2}+\bar{T}_{2}} -
  \frac{s_{\alpha}}{4\pi} \right\} \right] 
 + {\cal O} \left[ \frac{\delta_{GS}}{(T_2+\overline{T_2})^2} \right]
    \nonumber
\end{eqnarray}
where for simplicity we have chosen the $s_{\alpha}$ parameters to be the
1-loop beta-function coefficients (i.e. $s_{\alpha} = 2 \pi ( 33/5, 1, -3 )$
for $\alpha = U(1)_Y \, , \, SU(2)_L, \, , \, SU(3)_C$) in agreement
with Refs~\cite{ben,ben2}.  This can happen in ${\mathcal Z}_3$
and ${\mathcal Z}_7$ orbifold models, but in general $s_{\alpha}$ are
highly model dependent parameters related to the Green-Schwarz
coefficients~\cite{gs}. 
 
%\vskip5mm
\newpage
\noindent
$\bullet$ {\bf Soft trilinears:}
\vskip2mm

Recall that the allocation of MSSM states in
Eq.(\ref{eq:openall}) is constrained by the string selection rules in
Eq.(\ref{eq:wren}) to give a hierarchical Yukawa (and trilinear) texture with a
dominant (33) entry at leading-order.  The other elements are assumed
to arise from (unspecified) higher-dimensional operators in order to match the
measured first and second family fermion masses at the weak-scale
after RGE running.  Therefore we expect that (generically) the Yukawa
matrices have non-diagonal elements even at the high-scale.  However
for simplicity we choose to impose GUT-scale boundary conditions on
the trilinear couplings $A^f_{ij} \, Y^f_{ij}$ ($f=u,d,e$) in
Eq.(\ref{eq:lsoft}) that ignore the effect of any higher-dimensional
operators.  Then the GUT-scale trilinears have a single
non-zero (33) element:
\begin{eqnarray}
 A^f_{ij} \, Y^f_{ij}  = \left(
  \begin{array}{ccc}
   0 & 0 & 0 \\
   0 & 0 & 0 \\
   0 & 0 & {\cal A} \, Y^f_{33}
  \end{array} \right)  \hspace*{1cm} (f = u, d, e)
    \label{eq:tribc}
\end{eqnarray}
where $Y^f_{33}$ is the (33) element of the (running) Yukawa matrix
$Y^f$ at the GUT-scale.  Notice that we are not explicitly imposing a
hierarchical (33) texture on the GUT-scale Yukawa matrices, since
their form at the GUT-scale is determined by running the RGEs up from
the correct weak-scale values (after CKM mixing).  The universal soft
trilinear ${\cal A}$ is found to be:
\begin{eqnarray}
 {\cal A} \equiv 
  A_{C^{5_2}_1 C^{5_2}_2 C^{5_2}_3} = - m_{3/2} \cos\theta \, \sin\phi
   + {\cal O} \left[ \frac{\delta_{GS}}{(T_2 + \overline{T_2})^2} \right]
    \label{eq:softtri}
\end{eqnarray}
where we have assumed no explicit dilaton/moduli dependence in the
Yukawa couplings~\footnote{This is not entirely true in type I models
since the Yukawa couplings are equal to the D-brane gauge couplings
which are functions of the dilaton and moduli fields.  See
Refs.\cite{abel,ross,peddie2} for a recent discussion.}.

%%%%%%%%%%%%%%%%%%%%%%%%%%%%%%%%%%%%%%%%%%%%%%%%%%%%%%%%%
\subsubsection{Choice of model parameters}  \label{sec:choose}

At this stage we have a variety of (arbitrary) independent
parameters, and in order to make progress we make the
theoretically-motivated assumptions given in Table \ref{tab:paramcon}
to leave four independent parameters $m_{3/2}$, $\tan\beta$,
$\phi$ and $\theta$ to scan over.  Note that the $\theta$ Goldstino
angle is constrained to the range $0 \le \theta \le \arcsin(1/\sqrt{3})$ 
to avoid tachyonic soft Higgs doublet masses at the GUT-scale as
shown by Eq.(\ref{eq:softc22}).  This constraint allows the
electroweak symmetry to be radiatively broken by running the RGEs down to the
weak-scale instead of a tree-level breaking at the GUT-scale.
\TABLE{
 \scalebox{0.9}{
 \begin{tabular}{|c|c|c|} \hline
   Parameter & Constraint & Comments \\ \hline
   $^{\dagger} \,$ gravitino mass $m_{3/2}$ & 100 - 10,000 \, GeV &
  sets SUSY breaking scale \\
   $^{\dagger} \,$ Goldstino angle $\theta$ & 
     $0 \le \theta \le \arcsin(1/\sqrt{3})$ &
     for Higgs masses $m_{C^{5_2}_2}^2 \ge 0$ \\
   $^{\dagger} \,$ Goldstino angle $\phi$ & $0 \le \phi \le \pi/2$ & \\
   Goldstino angles $\Theta_{1,2,3}$ & $1/\sqrt{3}$ & \\\
   \kahler potential $\hat{K}(X)$ & 
     $\partial_X \hat{K} = 0$, \, $\partial_{X}^{2} \hat{K} = 1$ &
     vanishing F.I. term \\
   Green-Schwarz coefficient $\delta_{GS}$ & -10 & agrees with
     \cite{ben,ben2} \\   
   VEV of ($T_2 + \bar{T}_2$) & 50 & from gauge unification \\
   VEV of ($Y_2 + \bar{Y}_2$) & 0 & from gauge unification \\ \hline
   $^{\dagger} \,$ tan$\beta$ & 2-50 & \\
   $M_{GUT}$ & & determined by RGE running \\
   Sign of $\mu$-parameter & $sgn(\mu) > 0$ & motivated by $b \rightarrow s \,
   \gamma$, $g_{\mu}-2$ \\ \hline 
\end{tabular}}
\caption{The choice of high-scale ({\it upper}) and
low-scale ({\it lower}) parameters for our model, where those
marked with $\dagger$ are input parameters.}
\label{tab:paramcon} 
}

In section \ref{sec:sigma} we commented that the twisted moduli
\kahler potential is an unknown function of $Y_2, T_2$ and
$\delta_{GS}$ with an argument $X = Y_2 + \overline{Y_2}  
 - \delta_{GS} \, {\mathrm ln} ( T_2 + \overline{T_2} )$ required to cancel
modular anomalies~\cite{gs}.  We have chosen to follow previous
analyses~\cite{ben,ben2} and assume that the derivatives of
$\hat{K}(X)$ satisfy the following constraints: $\hat{K}' \equiv
\partial_X \hat{K}(X) = 0$ and $\hat{K}'' \equiv \partial_{X}^{2}
\hat{K}(X) = 1$, where $\partial_X \hat{K} = 0$ ensures that potentially
problematic Fayet-Illiopoulos terms $\xi_{FI} \sim \hat{K}'$ vanish.  
The Green-Schwarz parameter is a model-dependent negative integer
${\cal O}(-10)$, and for simplicity we set $\delta_{GS}=-10$.

In Appendix \ref{app:softterms} we discuss how to diagonalise the
\kahler metric when it is non-diagonal due to the mixing between $T_2$
and $Y_2$ induced by $\hat{K}(X)$.  We can define the SUSY
breaking F-terms, and therefore the soft parameters, by expanding in
inverse-powers of the $T_2+\overline{T_2}$ VEV.  This expansion is
reliable when $T_2+\overline{T_2}$ is sufficiently large in
string-scale $M_{\ast}$ units.  We also know that the (tree-level)
gauge couplings $g_{\alpha}$ are functions of $T_{2}+\overline{T_2}$ and 
$Y_{2}+\overline{Y_2}$ through the gauge kinetic functions 
$f_{\alpha} = T_2 + \frac{s_{\alpha}}{4\pi} Y_2$ and the relation 
${\mathrm Re} f_{\alpha} = 4\pi/g_{\alpha}^2$.  However the parameters
$s_{\alpha}$ are not equal for all three groups of the MSSM, which
implies that at tree-level the gauge couplings do not unify to a
single coupling $g_{GUT}^2 \approx 4\pi/25$.  Thus in order to
achieve unification we must appeal to (unspecified) higher-order
corrections to the gauge kinetic functions.  In this work we will
assume that these higher-order corrections only account for a small
correction to the tree-level value and so we choose that:
\begin{eqnarray}
 (T_2+\overline{T_2}) = 50 \hspace*{1cm} {\mathrm and} \hspace*{1cm} 
   (Y_2+\overline{Y_2}) = 0 
\end{eqnarray}
which is entirely consistent with coupling unification and different
$s_{\alpha}$.  This value of $T_{2}+\overline{T_2}$ is also
sufficiently large to justify the series expansion of F-terms, and
we can neglect the higher-order corrections to the soft parameters
in Eqs.(\ref{eq:softci},\ref{eq:softga},\ref{eq:softtri}).

Throughout our analysis we will neglect CP-phases and consequently
keep all soft scalar squared-masses positive-definite.  We demand a
neutral lightest supersymmetric particle (LSP), and
also that the low-energy gluino mass is not much larger than 1-2 TeV from
fine-tuning arguments~\cite{finetune}.  In section \ref{sec:fcnc} we
will compare generalised bounds on mass-insertion deltas derived from
CP-conserving FCNC experimental data~\cite{fcnc} to check that our
model is allowed.  We will take the Higgsino $\mu$-parameter to be
positive which is consistent with the $b\longrightarrow s \gamma$
observations and recent favoured values of $g_{\mu}-2$.
\TABLE{
 \scalebox{0.9}{
 \begin{tabular}{|c|c||c|c|} \hline
   Sparticle & Lower bound (GeV) & Sparticle & Lower bound (GeV) \\ \hline
    lightest Higgs \, $H_{1}^{0}$ & 111.4 
      & sleptons $\tilde{e}, \tilde{\mu}$ & 88 \\
    neutralino $\chi_{1}^{0}$ & 37 
      & stop $\tilde{t}_1$ & 86.4 \\
    chargino $\chi_{1}^{\pm}$ & 67.7 
      & sbottom $\tilde{b}_1$ & 91 \\
    gluino $\tilde{g}$ & 195 
      & squarks $\tilde{u}, \tilde{d}, \tilde{c}, \tilde{s}$ & 250 \\
    stau $\tilde{\tau}_1$ & 76 
      & sneutrino $\tilde{\nu}_1$ & 43.1 \\ \hline
\end{tabular}}
\caption{Experimental lower bounds on sparticle masses from
  direct searches~\cite{pdg}.  The first column of experimental limits provide
  the strongest constraints, where the MSSM-like lightest Higgs mass
  includes a 3 GeV theoretical uncertainty.}
\label{tab:sparticlecon} 
}

In addition to these theoretical constraints, we use the experimental 
limits on sparticle masses given in Table \ref{tab:sparticlecon} to
constrain our allowed parameter space.  Note that the experimental limit
for the MSSM-like lightest Higgs $h^0$ is $m_{h^0} = 114.4$ GeV.  We
use $111.4$ GeV, since we include a $\pm 3$ GeV error on SOFTSUSY's
prediction of $m_{h^0} = 114.4$ GeV.  We will comment on the
sparticle(s) that provide the strongest constraint to the viable
parameter space for a variety of model points.

We use the default version of SOFTSUSY 1.7.2~\cite{softsusy}, which uses 
2-loop RGEs for all parameters except sfermion masses and soft trilinears. 
Pole values for sparticle masses are calculated with one (and in sensitive 
cases) two-loop threshold contributions. We refer the reader to the 
SOFTSUSY manual~\cite{softsusy} for more details.
Electroweak gauge unification fixes the scale $M_X$ at which the boundary 
conditions on SUSY breaking masses are fixed:
\begin{equation}
 g_1(M_X) = g_2(M_X)  
   \label{eq:gcu}
\end{equation}
For $g_3$, we assume unknown high-scale corrections bring the prediction 
$g_3(M_X)=g_1(M_X)$ into line with the input value of the Standard Model 
$\alpha_s(M_Z)^{\overline MS}=0.1172$, which is used for our analysis.
We also use $m_t=174.3$ GeV, $m_b(m_b)^{\overline MS}=4.25$ GeV, the 
SOFTSUSY defaults. 

For the analysis of flavour changing neutral currents, a detailed Yukawa 
texture is considered beyond the scope of the present paper. We therefore 
allow low energy data on fermion masses and mixings to set the 
electroweak-scale Yukawa couplings (with an additional assumption about 
whether quark mixing resides entirely in the down, or up quark sector). We 
neglect neutrino masses and mixings, which will have an unimportant effect 
on the rest of the sparticle spectrum. We have added the boundary 
conditions in Eqs.(\ref{eq:ssms}-\ref{eq:softtri}) to the SOFTSUSY code.

In our analysis we consider also 
a slightly modified model from Eq.(\ref{eq:openall}) in which the superfields
$D^c_{3R}, E^c_{3R}$ are reassigned to $C_1^{5_2}$. String
selection rules will then forbid a renormalisable bottom and tau Yukawa 
coupling, allowing low $\tan\beta$
without changing the results for the soft masses
according to Eq.(\ref{eq:softc2j}).

%%%%%%%%%%%%%%%%%%%%%%%%%%%%%%%%%%%%%%%%%%%%%%%%%%%%%%%%%%
%%%%%%%%%%%%%%%%%%%%%%%%%%%%%%%%%%%%%%%%%%%%%%%%%%%%%%%%%%
%%%%%%%%%%%%%%% SECTION: RESULTS %%%%%%%%%%%%%%%%%%%%%%%%%
%%%%%%%%%%%%%%%%%%%%%%%%%%%%%%%%%%%%%%%%%%%%%%%%%%%%%%%%%%
%\newpage
\section{Results}  \label{sec:results}

In this section we will study the phenomenology of different
SI$\tilde{g}$M model 
points around the twisted moduli domination limit ($\theta=\phi=0$).
We observe that the viable parameter space is greatly increased by
including contributions to the SUSY breaking from gravity mediation
($F_S$ and/or $F_{T_i} \neq 0$) by increasing $\theta$, $\phi$.  In
contrast we will consider the heavily constrained untwisted T-moduli
domination limit ($\theta=0$, $\phi=\pi/2$)
and study the effect of including twisted moduli by reducing $\phi$.
We also study a couple of intermediate points with both $\theta$ and
$\phi$ non-zero, and perform some scans over $\theta$--$\phi$ with
fixed $m_{3/2}$ and $\tan\beta$.  We will discuss the implications of
FCNC bounds on our models, and give sample sparticle spectra for some
``benchmark'' points.

%%%%%%%%%%%%%%%%%%%%%%%%%%%%%%%%%%%%%%%%%
%%%%%%%%%%%%%% t0p0 %%%%%%%%%%%%%%%%%%%%%
%\newpage
\subsection{Twisted moduli domination: $\theta = \phi = 0$ } 
 \label{sec:t0p0}

We begin our phenomenological analysis with the twisted moduli
domination limit which has not been previously studied in the
literature.  In this limit the first and second family soft scalar
masses are exponentially small in comparison to the third family and Higgs
soft masses, and the dominant soft trilinear vanishes as shown in
the first column of Table \ref{tab:moveup}.  As a result the viable
parameter space is restricted to a very narrow range of $\tan\beta$
values, although there is still significant variation in the allowed
physical masses of the (a) lightest Higgs, (b) gluino, (c) neutralino
and (d) first/second family squarks as shown in Figure \ref{fig:t0p0}.  The
different treatment of the three families is highlighted in Figure
\ref{fig:t0p0comp} where the first/second family squarks are 
significantly lighter than the lightest stop states as a result of the
hierarchy of soft masses at the GUT-scale.  This leads to a
characteristic sparticle spectrum which we have labelled ``stop-heavy
MSSM''~\cite{bmsb}.  This constrasts with the usual minimal SUGRA case
where one stop is lighter than the other squarks due to top-Yukawa RGE
and mixing effects.
%\vskip-10mm
\FIGURE[h]{
 \epsfig{file=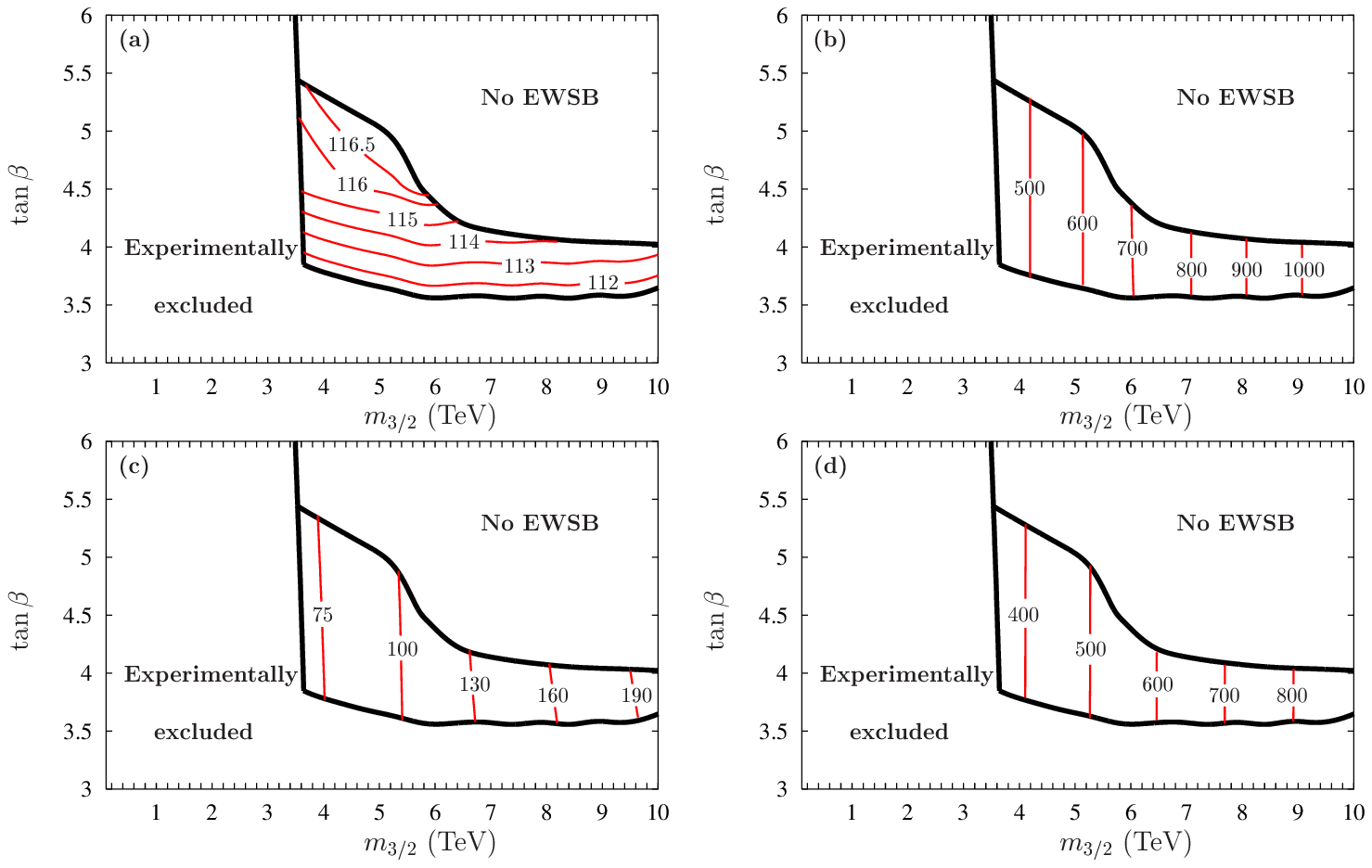,height=9cm}
 \caption{Maps of viable parameter space and mass contours of
(a) lightest Higgs $h^0$, (b) gluino $\tilde{g}$, (c) neutralino 
$\chi_{1}^{0}$ and (d) lightest squark $\tilde{q} = \tilde{u},
\tilde{d}, \tilde{c}, \tilde{s}$ for $\theta = \phi =0$.}  
 \label{fig:t0p0}
}

%\vskip-10mm
\TABLE{
 \scalebox{0.9}{
\begin{tabular}{|c||c|c|c|c|} \hline
   Soft mass/trilinear & (a) $\phi = 0$ & (b) $\phi = 0.05$ & 
     (c) $\phi = 0.07$ & (d) $\phi = 0.1$ \\
   (in $m_{3/2}$ units) & & & & \\ \hline
   1st/2nd family $C^{5_1 5_2}$ & 0.002 & 0.572 & 0.801 & 0.143 \\
   3rd family $C^{5_2}_{1,3}$ & 1 & 0.999 & 0.998 & 0.995 \\
   Higgses $C^{5_2}_{2}$        & 1 & 1 & 1 & 1 \\
   Trilinear $A_{C^{5_2}_1 C^{5_2}_2 C^{5_2}_3}$ 
                & 0 & -0.050 & -0.070 & -0.100  \\
   $M_1$ & 0.121  & 0.138 & 0.145 & 0.155 \\
   $M_2$ & 0.024  & 0.069 & 0.087 & 0.114 \\
   $M_3$ & -0.045 & 0.020 & 0.046 & 0.085 \\ \hline
\end{tabular}}
\caption{Numerical values of soft SUSY breaking parameters
  (in units of $m_{3/2}$) as $\phi$ is increased with $\theta=0$ kept
  fixed.  We assume that the unified GUT-scale
  gauge coupling is $\alpha_{GUT} \approx 1/25$ in order to estimate the
  relative size of the gaugino masses.}
\label{tab:moveup}
}

Low values of $m_{3/2} \le 3.5$ TeV are forbidden by the
chargino $\chi_{1}^{\pm}$ 
experimental bound, whereas the Higgs bound $m_{h^{0}} > 111$ GeV
rules out very small values of $\tan\beta \simlt 4$.  The other
constraint comes from the requirement of valid radiative EWSB which rules out
the space for $\tan\beta > 4.5$ and $m_{3/2} > 3.5$ TeV.

\FIGURE[h]{
 \epsfig{file=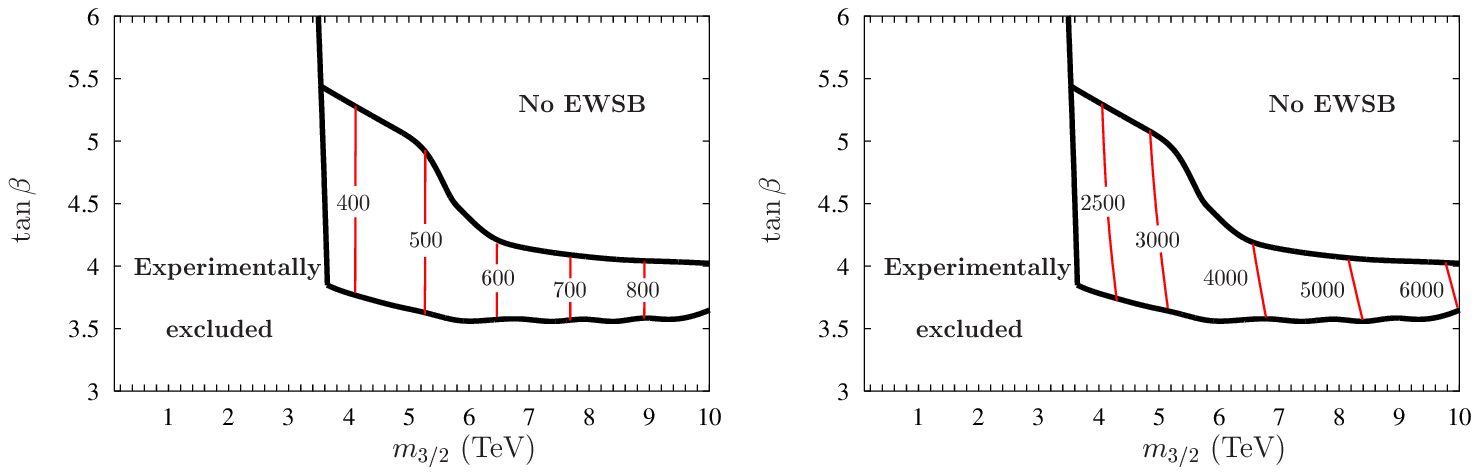,height=4.7cm}
 \caption{Comparison of the mass contours for 
({\it left}) first/second family squarks $\tilde{q} = \tilde{u},
\tilde{d}, \tilde{c}, \tilde{s}$ and ({\it right}) stop squark
$\tilde{t}_1$ for $\theta = \phi =0$.}
 \label{fig:t0p0comp}
}

There is an intriguing experimental signature of the twisted moduli
limit from the quasi-degeneracy of the lightest neutralino and
chargino states.  This is similar to the situation in
anomaly mediated SUSY breaking~\cite{amsb}, and can be understood from
the first column of 
Table \ref{tab:moveup} since the soft bino and wino gaugino masses
have the ratio $M_1 \approx 5 \, M_2$.  Therefore the lightest chargino mass
eigenstate ($m_{\chi_{1}^{\pm}}$) is roughly the wino ($M_2$), and so is
the lightest neutralino mass eigenstate ($m_{\chi_{1}^{0}}$) (up to
electroweak corrections).  The light Higgs mass $m_{h^0} < 117$ GeV
and heavy stop are other indicators of the scenario.

%%%%%%%%%%%%%%%%%%%%%%%%%%%%%%%%%%%%%%%%%%%%%%%%%%%%%%%%%%%%%%%%
%%%%% compare increasing phi away from twisted Y-domination %%%%
%\newpage
\subsection{Moving away from twisted moduli domination by switching on
      T-moduli}  \label{sec:t0pinc}

In this section we will try to understand the effect of the twisted
moduli F-term on the viable parameter space by {\it increasing} the value of
$\phi$ to move away from the twisted moduli domination limit.
We keep $\theta=0$ fixed which eliminates any contributions from the
dilaton ($F_S = 0$), and so increasing $\phi$ includes the
untwisted T-moduli F-term contributions to the overall SUSY breaking VEV.

In Table \ref{tab:moveup} we show how the numerical values of the
GUT-scale soft parameters (in $m_{3/2}$ units) are affected by
increasing $\phi$.   The common first and second family soft scalar
masses rapidly increase, while the third family and Higgses remain
more or less constant.  The soft trilinear is no longer vanishing, and
the soft gaugino masses also rise - in fact the gluino soft mass changes
sign relative to the twisted moduli domination limit.  
%\vskip-10mm
\FIGURE[h]{
 \epsfig{file=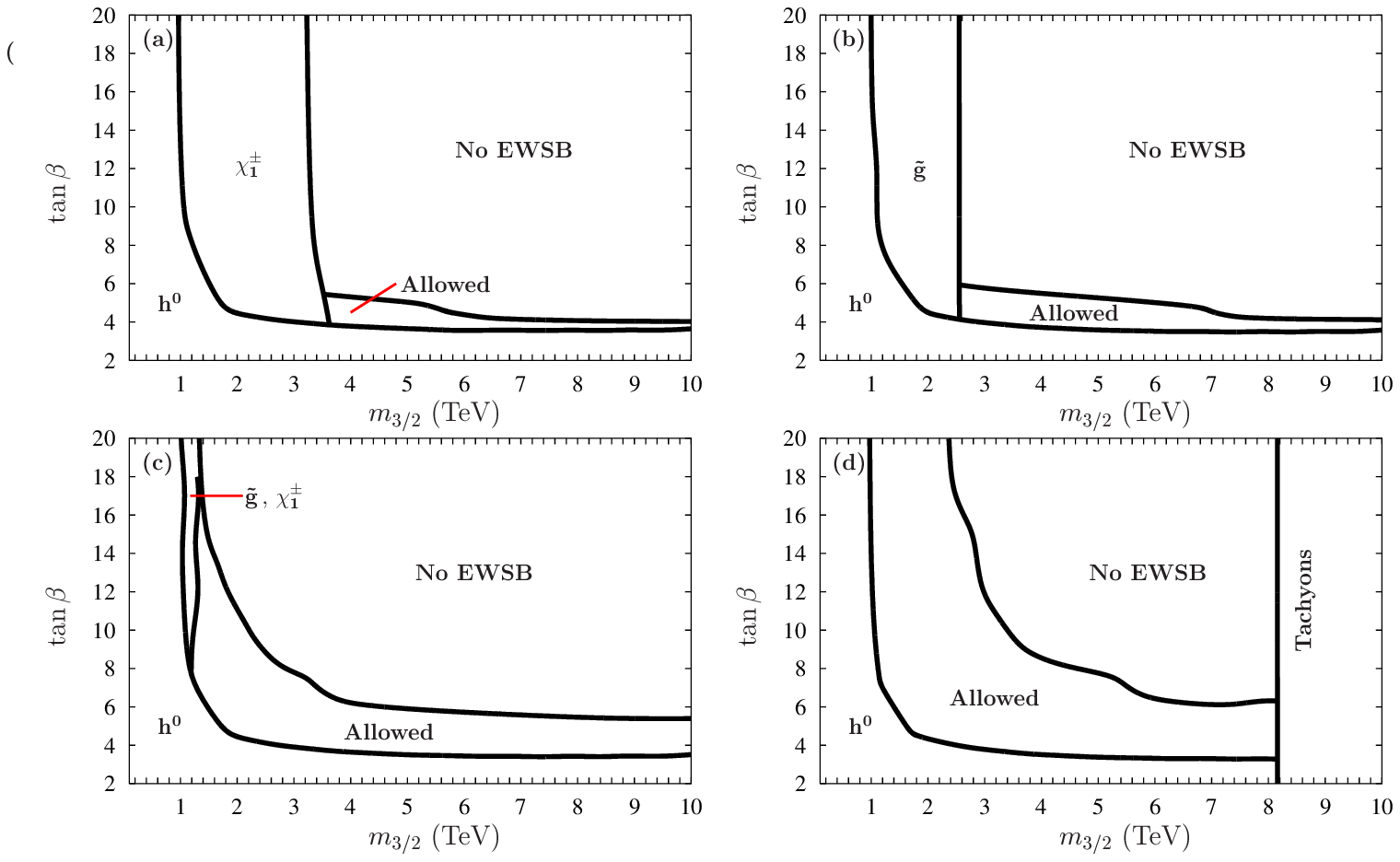,height=9cm}
 \caption{The viable parameter space ({\it Allowed}) opens up
as $\phi$ is increased with $\theta =0$ fixed; (a) $\phi=0$, (b)
$\phi=0.05$, (c) $\phi=0.07$ and (d) $\phi=0.1$.  Note that (d)
extends up to include $\tan\beta=50$.  We label the surrounding
regions of parameter space by the sparticle(s) that provide the strongest
constraints.}  
 \label{fig:moveup}
}
We plot the evolution of the viable parameter space in Figure
\ref{fig:moveup}, and as expected we observe that the space rapidly
opens up with increasing $\phi$.  In each case the lightest Higgs mass
experimental limit excludes very small $\tan\beta$ regions.  Moving
from (a) to (d) we see that the strongest constraints on the lower
$m_{3/2}$ region of parameter space comes from the chargino, gluino
and finally lightest Higgs experimental bounds. The remaining forbidden
space is ruled out by the requirement of EWSB.  It is interesting to
observe that for larger values of $\phi \simgt 0.1$ the
large-$m_{3/2}$ region is excluded by tachyons, and this 
tachyon boundary sweeps towards small-$m_{3/2}$ as $\phi$ increases
further.  Although not shown, the inclusion of non-zero $F_{T_i}$ also
lifts the degeneracy of the lightest neutralino and chargino states.

%%%%%%%%%%%%%%%%%%%%%%%%%%%%%%%%%%%%%%%%%%%%%%%%%%%%%%%%%%%%%%%%%%%%%
%%%%%%%% compare decreasing phi away from untwisted T-domination %%%%
\subsection{Moving away from untwisted T-moduli domination by
switching on twisted moduli}  \label{sec:t0pdec} 
 
In contrast to section \ref{sec:t0pinc} we will now focus on the other
extreme case of untwisted T-moduli domination, and study the effect of
including the twisted moduli contributions by {\it decreasing} $\phi$
away from its maximal $\pi/2$ value.  Once more we keep $\theta=0$
fixed to eliminate the dilaton effects.  Table \ref{tab:movedown}
gives the numerical values of the soft GUT-scale parameters in
units of $m_{3/2}$.  Note that the Higgs doublet soft mass remains fixed,
while the first and second family mass is also approximately
constant.  The third family scalar mass vanishes initially in the limit of
untwisted moduli domination, while the wino
and gluino (bino) soft gaugino masses decrease (increase)
respectively.
%\vskip-3mm
\TABLE{
 \scalebox{0.9}{
\begin{tabular}{|c||c|c|c|c|} \hline
   Soft mass/trilinear & (a) $\phi = \pi/2$ & (b) $\phi = 15 \pi/32$ & 
     (c) $\phi = 7\pi/16$ & (d) $\phi = 3\pi/8$ \\
   (in $m_{3/2}$ units) & & & & \\ \hline
   1st/2nd family $C^{5_1 5_2}$ & 11.45 & 11.39 & 11.23 & 10.58 \\
   3rd family $C^{5_2}_{1,3}$ & 0 & 0.098 & 0.195 & 0.383 \\
   Higgses $C^{5_2}_{2}$        & 1 & 1 & 1 & 1 \\
   Trilinear $A_{C^{5_2}_1 C^{5_2}_2 C^{5_2}_3}$ 
                & -1 & -0.995 & -0.981 & -0.924  \\
   $M_1$ & 0.340 & 0.350 & 0.357 & 0.361 \\
   $M_2$ & 0.900 & 0.898 & 0.887 & 0.841 \\
   $M_3$ & 1.300 & 1.289 & 1.266 & 1.184 \\ \hline
\end{tabular}}
\caption{Numerical value of soft SUSY breaking parameters
   (in units of $m_{3/2}$ with $\alpha_{GUT} \approx 1/25$) as $\phi$
   is decreased with $\theta=0$ kept fixed.}
\label{tab:movedown}
}

%\vskip-10mm
\FIGURE[h]{
 \epsfig{file=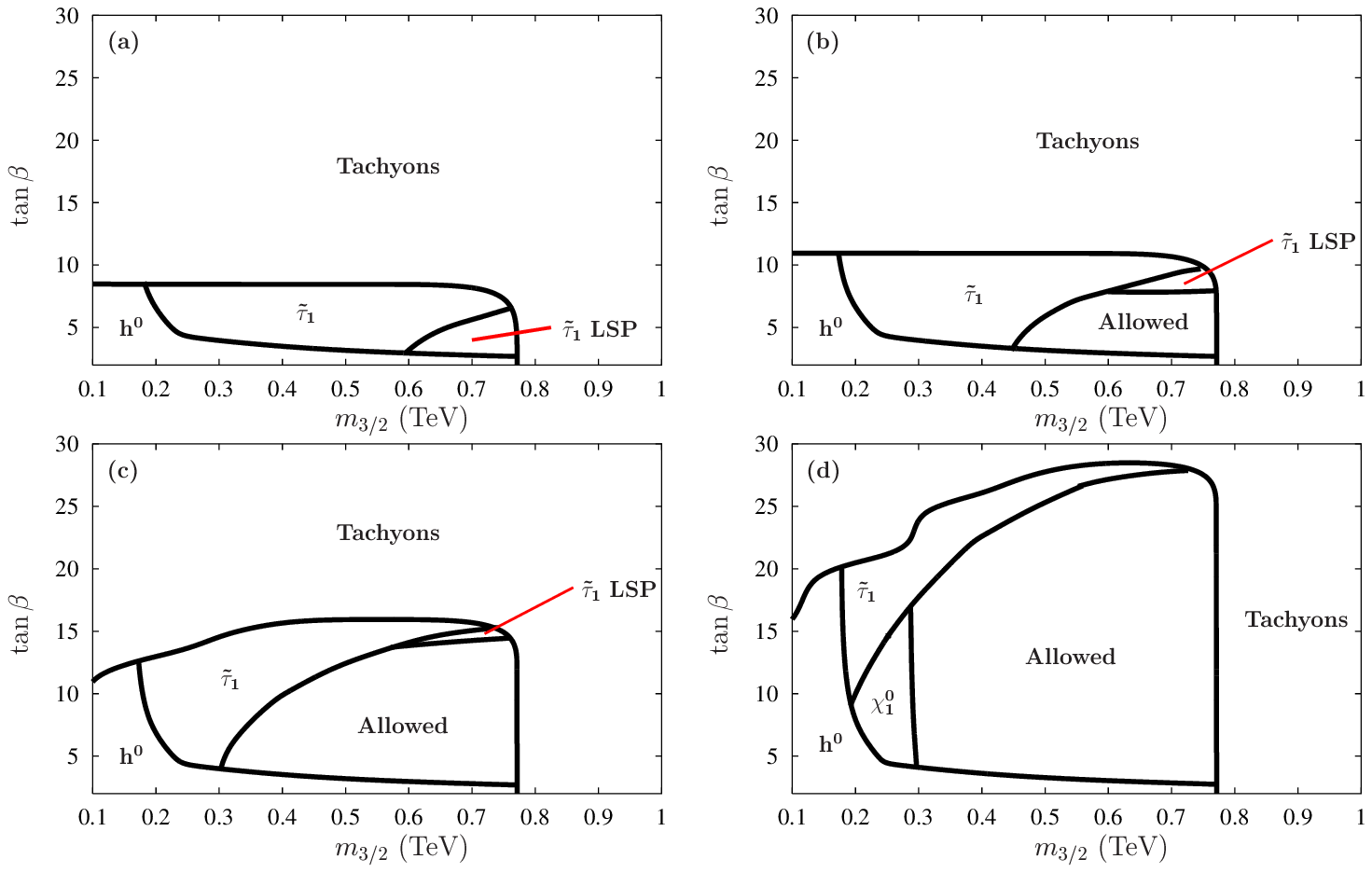,height=9cm}
 \caption{The viable parameter space ({\it Allowed}) opens up
as $\phi$ is increased with $\theta =0$ fixed; (a) $\phi=\pi/2$, (b)
$\phi=15\pi/32$, (c) $\phi=7\pi/16$ and (d) $\phi=3\pi/8$.  The viable
space for (a) has $\tilde{\tau}_1$ as the LSP and is therefore excluded
anyway.  We label the surrounding regions of parameter space with the
sparticle(s) that provide the strongest constraints.} 
 \label{fig:movedown}
}

Figure \ref{fig:movedown} shows the evolution of the parameter space
as we move away from T-moduli domination.  The first important feature
to notice from (a) is that only a very tiny patch of parameter space
is allowed since the majority is forbidden by low-energy tachyons.
The strongest constraints come from the lightest Higgs and stau (later
lightest neutralino) experimental bounds, but the allowed region
for T-moduli domination has a stau LSP and is therefore ruled out
anyway by experimental constraints (e.g. searches for anomalously
heavy nuclei predicted by nucleosynthesis).  We observe from (a) to (d) how the
inclusion of twisted moduli opens up the viable region by
pushing back the stau/neutralino boundaries, lifting their degeneracy
and yielding a neutralino LSP.  Decreasing $\phi$ even
further also pushes back the tachyon boundary to larger $m_{3/2}$,
until we reach the limits considered in Section \ref{sec:t0pinc} where
the experimental sparticle constraints reappear at low $m_{3/2}$.

%%%%%%%%%%%%%%%%%%%%%%%%%%%%%%%%%%%%%%%%%%%%%%%%
%%%%%%%%%%%%%%%%%%%%%%%%%%%%%%%%%%%%%%%%%%%%%%%%
\subsection{General SUSY breaking with dilaton {\it and} moduli F-terms}
  \label{sec:general}
 
In the previous sections we have observed the effects of varying
$\phi$ (with $\theta=0$ fixed) on the evolution of the viable
parameter space.  We have seen that both twisted and untwisted moduli
dominated limits have severely constrained parameter spaces that can be
enlarged by including the effects from the other moduli F-term.
We can predict that the least constrained models are likely to
involve both types of moduli contributions (and also dilaton effects),
and in this section we will briefly study two sample points with
non-zero $\theta$, $\phi$ Goldstino angles.

%%%%%%%%%%%%%% t0_1p0_1 %%%%%%%%%%%%%%%%%%%%%%%%
\subsubsection{Twisted moduli domination with dilaton and moduli
F-terms: $\theta = \phi = 0.1$}   \label{sec:t0_1p0_1}

We will first study a point close to the twisted moduli domination
limit with $\theta=\phi=0.1$, where we include additional
contributions from the dilaton and untwisted moduli F-terms.  
The GUT-scale soft masses are given in Table
\ref{tab:t0_1p0_1}. 
%\vskip-3mm

%\vskip-10mm
\FIGURE[h]{
 \epsfig{file=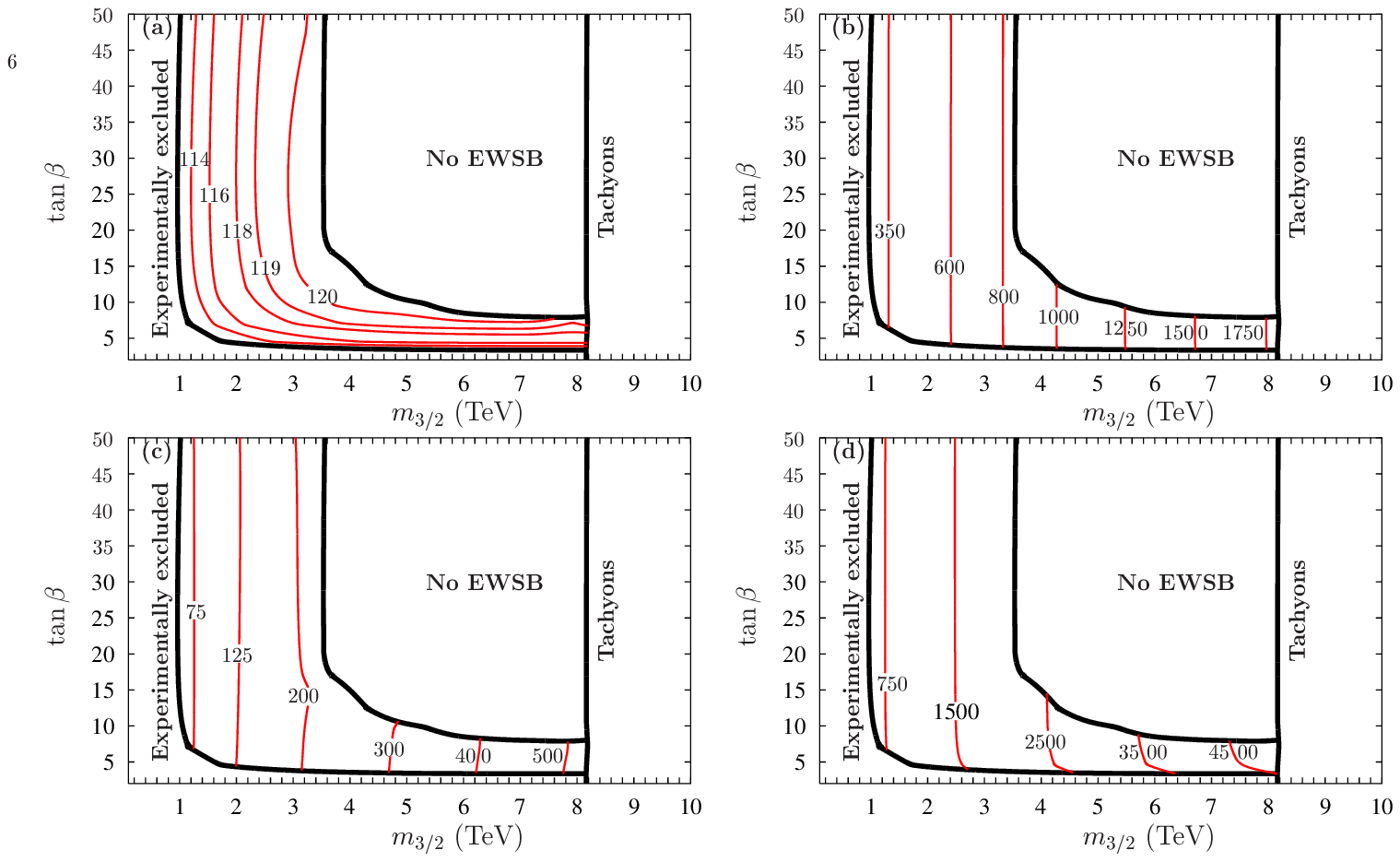,height=9cm}
 \caption{Maps of viable parameter space and mass contours of
(a) lightest Higgs $h^0$, (b) gluino $\tilde{g}$, (c) neutralino
$\chi_{1}^{0}$ and (d) lightest stop $\tilde{t}_1$ for 
$\theta = \phi =0.1$.}
 \label{fig:t0_1p0_1} 
}
\TABLE{
 \scalebox{0.9}{
\begin{tabular}{|c|c||c|c|} \hline
   Soft mass/trilinear & (in $m_{3/2}$ units) &
    Soft mass/trilinear & (in $m_{3/2}$ units) \\ \hline
   1st/2nd family $C^{5_1 5_2}$ & 1.135 & $M_1$ & 0.154 \\
   3rd family $C^{5_2}_{1,3}$ & 0.995 & $M_2$ & 0.113 \\
   Higgses $C^{5_2}_{2}$ & 0.985 & $M_3$ & 0.085 \\ 
   Trilinear $A_{C^{5_2}_1 C^{5_2}_2 C^{5_2}_3}$ & -0.0993 & & \\ \hline
\end{tabular}}
\caption{Numerical value of soft SUSY breaking parameters
   (in units of $m_{3/2}$ with $\alpha_{GUT} \approx 1/25$) for
   $\theta=\phi=0.1$.}
\label{tab:t0_1p0_1}
}

Comparing the results shown in Figure \ref{fig:t0_1p0_1} with
Figure \ref{fig:moveup} (d) we conclude that there
is only a very weak dependence on $\theta$.  In terms of constraints,
the main differences to the twisted moduli dominated limit are the
range of allowed $\tan\beta$ values and the presence of tachyons
at large $m_{3/2}$.  The lightest Higgs limit provides the only
experimental constraint at low $\tan\beta$ and small $m_{3/2}$.  This
model point allows heavier particles, for example the lightest Higgs
$h^0$ can be as heavy as $120$ GeV.

%%%%%%%%%%%%%%%%%%%%%%%%%%%%%%%%%%%%%%%%%%%%%%
%%%%%%%%%%%%%% t0_615p0_1 %%%%%%%%%%%%%%%%%%%%
\subsubsection{Vanishing soft Higgs mass: 
$\theta = \arcsin(1/\sqrt{3})$, $\phi = 0.1$}  \label{sec:t0_615p0_1}

We will now study an interesting model point where the Higgs doublet
soft mass vanishes at the GUT-scale by choosing
$\theta=\arcsin(1/\sqrt{3})$, as shown by Eq.(\ref{eq:softc22}).  We
set $\phi=0.1$ to avoid first/second
family ($\ci$) tachyonic soft masses.  The numerical values of the
corresponding soft parameters are given in Table \ref{tab:nohiggs}.
%\vskip-7mm
%\vskip-10mm
\FIGURE[h]{
 \epsfig{file=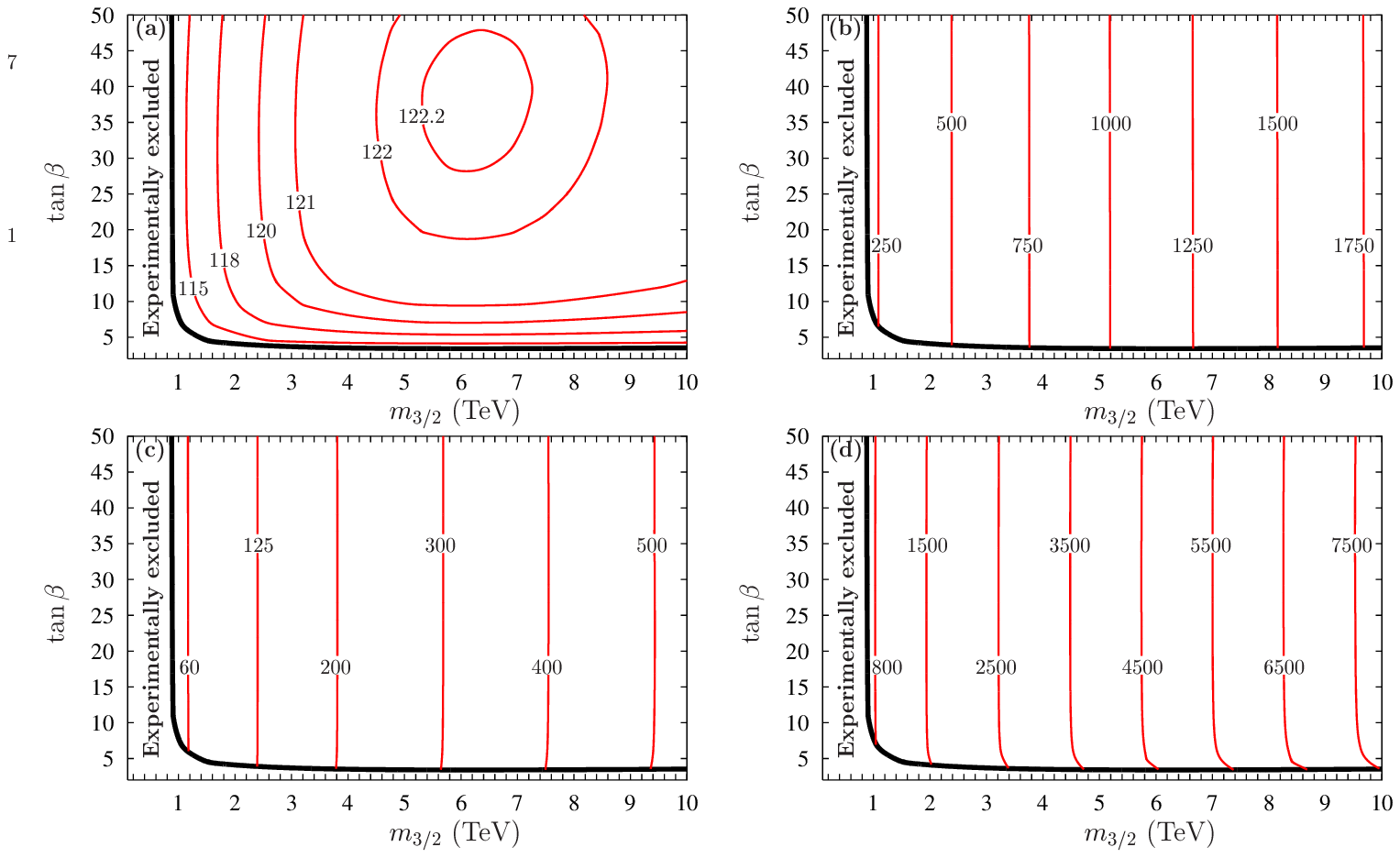,height=9cm}
 \caption{Maps of viable parameter space and mass contours of
(a) lightest Higgs $h^0$, (b) gluino $\tilde{g}$ and (c) neutralino
$\chi_{1}^{0}$ and (d) stop $\tilde{t}_1$ for 
$\theta = \arcsin(1/\sqrt{3})$ , $\phi =0.1$ where the soft Higgs
doublet mass vanishes at the GUT-scale.}
 \label{fig:nohiggs}
}
\TABLE{
 \scalebox{0.9}{
\begin{tabular}{|c|c||c|c|} \hline
   Soft mass/trilinear & (in $m_{3/2}$ units) &
    Soft mass/trilinear & (in $m_{3/2}$ units) \\ \hline
   1st/2nd family $C^{5_1 5_2}$ & 0.839 & $M_1$ & 0.126 \\
   3rd family $C^{5_2}_{1,3}$ & 0.997 & $M_2$ & 0.093 \\
   Higgses $C^{5_2}_{2}$ & 0 & $M_3$ & 0.069 \\ 
   Trilinear $A_{C^{5_2}_1 C^{5_2}_2 C^{5_2}_3}$ & -0.082 & & \\ \hline
\end{tabular}}
\caption{Numerical value of soft SUSY breaking parameters
   (in units of $m_{3/2}$ with $\alpha_{GUT} \approx 1/25$) for
   $\theta = \arcsin(1/\sqrt{3})$, $\phi=0.1$.} 
\label{tab:nohiggs}
}

The viable parameter space and sparticle mass contours are shown in
Figure \ref{fig:nohiggs}.  It is clear that the vanishing soft mass for
the Higgs doublets does not restrict the allowed space, and the only
constraint originates from the lightest Higgs (and also chargino)
experimental bounds at
low values of $\tan\beta$ and $m_{3/2}$.  The vanishing Higgs 
mass ensures that there are no regions ruled out by the EWSB
constraints, and also tachyonic scalars do not exclude regions with
large $m_{3/2}$.  The sparticle mass contours take 
similar values to the previously studied model points, except that the
lightest stop can now be much heavier, and the lightest Higgs can reach
$m_{h^0} \approx 122$ GeV.

%%%%%%%%%%%%%%%%%%%%%%%%%%%%%%%%%%%%%%%%%%%%
%%%%%%%%%%%%% FCNC %%%%%%%%%%%%%%%%%%%%%%%%%
\subsection{Comments on Flavour Changing Neutral Current
constraints}    \label{sec:fcnc}

One of the most appealing features of sequestered
models is the geometrical explanation for the suppression of
flavour changing processes due to negligible direct
tree-level couplings between Standard Model and SUSY breaking fields.
Using EFT techniques, we find that the contributions from
non-renormalisable higher dimensional operators are
exponentially-suppressed by the separation $r$  between branes and the
UV cutoff scale $\Lambda_{UV}$ after integrating out the high-scale
physics.  In fact the dominant contributions to flavour changing
off-diagonal mass-matrix elements in these parallel-brane models arise
from loop corrections 
involving bulk fields.  Therefore the squarks and sleptons
start out almost massless at the GUT-scale, and receive soft masses at
low energy only through RGE running effects involving the gauge
couplings which couple in a flavour-blind way, 
leading to universal soft mass matrices at the low
scale.  The same mechanism suppresses the first/second
family FCNCs in the twisted moduli domination limit of our SI$\tilde{g}$M
model.  The separation of the third family fields with a heavier soft
scalar mass is allowed since the third family FCNCs are only weakly
constrained by experiment.  In fact a heavier third family at
low-energies proves helpful in achieving successful EWSB and a
sufficiently heavy lightest CP-even Higgs mass.

It is common to parameterise the amount of flavour violation using the
mass-insertion (MI) approximation.  In this work we have
generalised the CP-conserving constraints on the MI
deltas~\footnote{We have assumed that all soft parameters are
automatically real by neglecting CP-phases in the F-terms, and hence
we cannot utilise the experimental electric dipole moment and
$\epsilon'/\epsilon$ bounds 
to constrain our models.  However we will not invoke the experimental
lepton flavour violation limits since we do not claim to have a theory
of neutrino masses in this work.} derived from experiment in
Ref.~\cite{fcnc} to test the viability of our model for various
$\theta$--$\phi$ points.  The results of comparing the predicted MI
deltas with the experimental limits for these points is shown in Table
\ref{tab:mid}. 
%\vskip-5mm
\TABLE{
\scalebox{0.8}{
\begin{tabular}{|c||c|c||c||c|} \hline
      Ratio $R[\delta] = \delta_{SS}/\delta_{exp}$ 
      & \multicolumn{2}{c||}{$\theta=\phi = 0$}  
      & $\theta=\phi = 0.1$ & $\theta=0.615, \, \phi = 0.1$ \\ \hline
      \vrule height 13pt width 0pt
     Weak-scale mixing & up & down & up & up \\ \hline
     $R[(\delta^{d}_{11})_{LR}]$ 
      & $(5.4 - 5.1) \times 10^{-3}$ 
      & $(5.4 - 5.1) \times 10^{-3}$ 
      & $(5 - 3.7) \times 10^{-4}$ 
      & $(6 - 4) \times 10^{-4}$  \\
     $R[(\delta^{d}_{22})_{LR}]$ 
      & $(6.3 - 6) \times 10^{-3}$ 
      & $(5.6 - 5.2) \times 10^{-3}$ 
      & $(6 - 4.4) \times 10^{-4}$ 
      & $(7 - 4.6) \times 10^{-4}$  \\
     $R[(\delta^{d}_{12})_{LL}]$ 
      & $(7 - 3.5) \times 10^{-2}$ 
      & $0.3 - 0.1$ 
      & $(1.6 - 0.4) \times 10^{-3}$ 
      & $(1.2 - 0.3) \times 10^{-3}$  \\
     $R[(\delta^{d}_{12})_{LR}]$ 
      & $(6 - 1.5) \times 10^{-7}$ 
      & $(6 - 1.5) \times 10^{-7}$ 
      & $(1 - 0.04) \times 10^{-8}$ 
      & $(1.5 - 0.05) \times 10^{-8}$  \\
     $R[\sqrt{(\delta^{d}_{12})_{LL} (\delta^{d}_{12})_{RR}}]$ 
      & $(2 - 0.8) \times 10^{-4}$ 
      & $6.5 - 3.5$ 
      & $(2 - 0.1) \times 10^{-6}$ 
      & $(5 - 0.2) \times 10^{-6}$  \\
     $R[(\delta^{u}_{12})_{LL}]$ 
      & $0.1 - 0.05$ 
      & $(3 - 1) \times 10^{-4}$ 
      & $(5 - 0.8) \times 10^{-4}$ 
      & $(5 - 0.5) \times 10^{-4}$  \\
     $R[(\delta^{u}_{12})_{LR}]$ 
      & $(8 - 1) \times 10^{-9}$ 
      & $(8 - 1) \times 10^{-9}$ 
      & $(3 - 0.15) \times 10^{-5}$ 
      & $(5 - 0.2) \times 10^{-5}$  \\
     $R[\sqrt{(\delta^{u}_{12})_{LL} (\delta^{u}_{12})_{RR}}]$ 
      & $0.8 - 0.45$ 
      & $(1 - 0.5) \times 10^{-7}$ 
      & $(7 - 1) \times 10^{-4}$ 
      & $(1 - 0.1) \times 10^{-3}$  \\
     $R[(\delta^{d}_{23})_{LL}]$ 
      & $(3 - 0.7) \times 10^{-3}$ 
      & $(8 - 2) \times 10^{-3}$ 
      & $(1 - 0.03) \times 10^{-3}$ 
      & $(1 - 0.02) \times 10^{-3}$  \\ 
     $R[(\delta^{d}_{23})_{LR}]$ 
      & $(2 - 0.3) \times 10^{-5}$ 
      & $(2 - 0.3) \times 10^{-5}$  
      & $(5 - 0.1) \times 10^{-5}$ 
      & $(1 - 0.01) \times 10^{-4}$  \\ \hline
\end{tabular}}
\caption{Ranges of the ratio $R[\delta] =
      \delta_{SS}/\delta_{exp}$ of mass-insertion deltas for the three
      different $\theta$--$\phi$ model points in sections
      \ref{sec:t0p0}, \ref{sec:t0_1p0_1} and \ref{sec:t0_615p0_1},
      with weak-scale mixing in either the up or down-quark sectors.
      The ranges come from scanning over allowed points in the
      $\tan\beta$--$m_{3/2}$ plane.}
\label{tab:mid}
}
Recall from Ref.~\cite{fcnc} that each CP-conserving flavour changing
observable is calculable in terms of combinations of MI 
deltas, e.g. $\Delta m_K$ from $K^0$--$\overline{K}^0$
mixing and BR($b\longrightarrow s \, \gamma$).  Assuming no
accidental cancellations we can use experimental bounds to
calculate upper limits on each separate MI delta as functions of
the physical squark and gluino masses. For every
$\tan\beta$--$m_{3/2}$
point within the allowed parameter space, we can calculate the ratio 
$R[\delta] = \delta_{SS} / \delta_{exp}$ between the actual
MI delta matrix element $\delta_{SS}$ (calculated by SOFTSUSY) and
the corresponding experimental limit $\delta_{exp}$, where regions
with a ratio greater that unity are ruled out.  
In Figure \ref{fig:t0p0deltas} we plot contours of the
ratio $R[\delta]$ for two particular delta matrix elements with
weak-scale mixing in the up-quark sector - $(\delta^{u})_{LL}$ and  
$\sqrt{(\delta^{u}_{12})_{LL} \, (\delta^{u}_{12})_{RR}}$ - in the
limit of twisted moduli domination, that are very close to being excluded
by experiment.  We obtain similar plots for the other MI deltas, and
also for the $\theta=\phi=0.1$ and $\theta=0.615, \phi=0.1$ model
points, and the relative ratios $R[\delta]$ are compared in Table
\ref{tab:mid}.   We also include the ratios for the twisted moduli
domination limit, but with weak-scale mixing shifted to the down-quark sector.
%\vskip-10mm
\FIGURE[h]{
 \epsfig{file=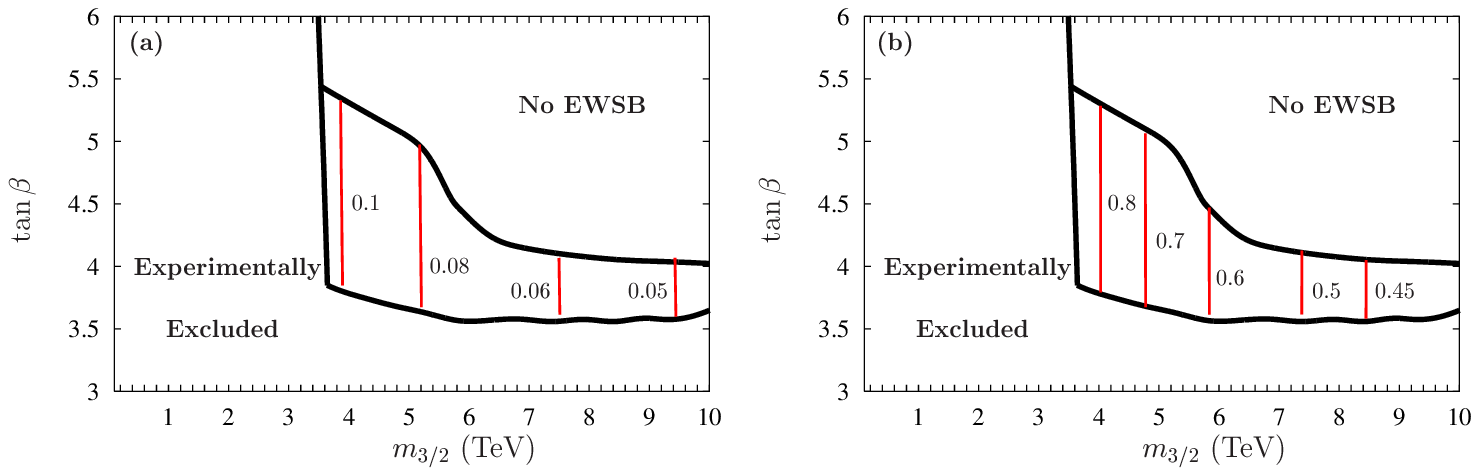,height=4.7cm}
 \caption{The ratio between mass-insertion deltas predicted by
SOFTSUSY (with weak-scale mixing in the up-quark sector only) 
against the bounds deduced from experiment~\cite{fcnc} 
for $\theta=\phi=0$: {\it (a)} $(\delta^{u}_{12})_{LL}$,
{\it (b)} $\sqrt{(\delta^{u}_{12})_{LL} (\delta^{u}_{12})_{RR}}$ are
closest to being ruled out or probed by experiment.}
 \label{fig:t0p0deltas}
}

We observe that the model points (with weak-scale mixing in the
up-quark sector) in Table \ref{tab:mid}
easily satisfy the experimental constraints with ranges of ratios 
$R[\delta] \ll 1$.  We have already discussed that the mechanism for
suppressing FCNCs in the twisted moduli limit is analogous to the
\gmsb models.  However there is a different mechanism at work that
solves the SUSY flavour problem when we move away from twisted moduli
domination since the inclusion of gravitational SUSY breaking effects
spoils the sequestering to give non-negligible first/second family
soft sfermion masses at the high-scale.  The 
suppression of FCNCs can be understood by studying the expressions for the soft
parameters in Eqs.(\ref{eq:ssms}-\ref{eq:softtri}) and observing how
the MI deltas vary with different gluino and first/second family squark
masses~\cite{fcnc}.
For example, the diagonal soft mass-squared $m_{\ci}^2$ for the
first/second family squarks grows rapidly with increasing $\phi$ since
the first term in the second line of Eq.(\ref{eq:softci}) dominates.
Also the gluino mass increases as the first term in
Eq.(\ref{eq:softga}) dominates, and a heavier gluino leads 
to heavier weak-scale squarks through RGE running.  The heavier
gluino and squark masses, in combination with the assumption of
family-diagonal soft scalar mass-matrices at the high-scale, are
sufficient to suppress the first/second family FCNCs and ameliorate
the SUSY flavour problem.

The values of the ratios $R[\delta]$ given in
Table~\ref{tab:mid} for the twisted moduli domination limit are shown
to have an explicit dependence on
whether the weak-scale boundary conditions for the quark Yukawa matrix
parameters are imposed in the up or down-quark
sectors, although the physical sparticle masses remain
unaffected.  For instance if the weak-scale mixing occurs 
exclusively in the down-quark sector, then the mixing induced in the
up-quark sector should be relatively small since the up Yukawa has
more effect on the up-squarks.  Comparing the relative
ratios $R[\delta]$ for the twisted moduli domination limit in Table
\ref{tab:mid}, we see that moving the weak-scale Yukawa mixing into the
down-quark sector has little effect on the mixed-parity MI deltas --
$(\delta^{d}_{11})_{LR}$, $(\delta^{2}_{22})_{LR}$,
$(\delta^{d}_{12})_{LR}$, $(\delta^{u}_{12})_{LR}$ and
$(\delta^{d}_{23})_{LR}$ -- but greatly affects the same-parity MI
deltas (although $(\delta^{d}_{23})_{LL}$ only slightly changes).  The
main effect is to weaken the constraints on $(\delta^{u}_{12})_{LL}$ and  
$\sqrt{(\delta^{u}_{12})_{LL} \, (\delta^{u}_{12})_{RR}}$, that are
close to experimental limits with up-quark sector mixing as shown in Figure
\ref{fig:t0p0deltas}, but strengthen the constraints on the
equivalent down sector MI deltas $(\delta^{d}_{12})_{LL}$ and  
$\sqrt{(\delta^{d}_{12})_{LL} \, (\delta^{d}_{12})_{RR}}$.  In fact,
the value for $\sqrt{(\delta^{d}_{12})_{LL} \,
(\delta^{d}_{12})_{RR}}$ implies that the twisted moduli limit is
ruled out by experiment, albeit with weak-scale mixing only in
the down-quark sector.

%%%%%%%%%%%%%%%%%%%%%%%%%%%%%%%%%%%%%%%%%%%%%%%%%%%%%%%%%%%%%%%%%%%%%
%%%%%%%% compare param spaces for angle scans %%%%%%%%%%%%%%%%%%%%%%%
\subsection{Parameter scans over $\theta$--$\phi$}  \label{sec:angscan}

In this section we will study the $\theta$--$\phi$ dependence of our
SI$\tilde{g}$M model for different choices of $\tan\beta$ and $m_{3/2}$.
The unitarity of the Goldstino angle parameterisation in
Eq.(\ref{eq:ftermdef}) - where 
$F_{S}^2 + \sum_i F_{T_i}^2 + F_{Y_2}^2 = {\mathbf F}^2$ - enables us
to plot points in the square $\theta$--$\phi$ parameter space as
points within a 
triangle~\footnote{Similar triangular plots arise when
discussing the relative mixing of eigenstates in neutrino 
oscillations~\cite{triangle}.} with sides labelled by $\sin^2(\theta)$
and $\sin^2(\phi)$, 
where each triangle vertex corresponds to a particular limit of SUSY
breaking as shown in Figure \ref{fig:tribasic}.

\FIGURE[h]{
 \epsfig{file=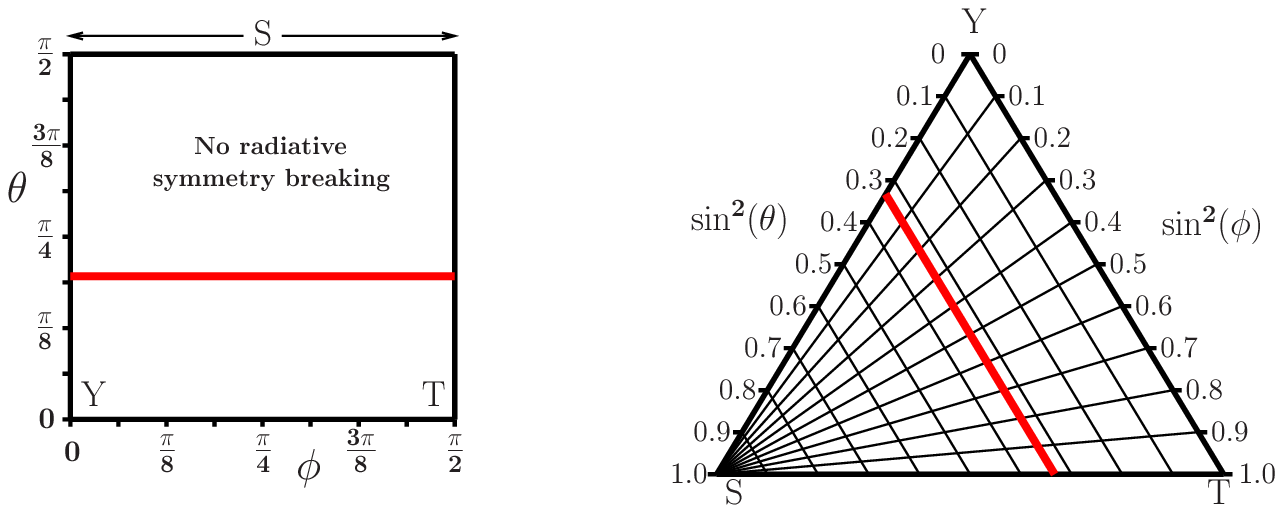,height=5.5cm}
 \caption{Generic points within the square $\theta$--$\phi$
parameter space ({\it left panel}) can be mapped on to a
triangular coordinate system $\sin^2(\theta)$--$\sin^2(\phi)$ 
({\it right panel}).  The region above/left of the thick (red) line is
forbidden since there is no radiative EWSB as the 
symmetry is already broken at the GUT-scale by tachyonic Higgs
doublet soft scalar masses.  We also label the different limits of SUSY
breaking domination ($S, T, Y$) which correspond to the vertices of
the triangle.} 
 \label{fig:tribasic}
}

%\vskip-15mm
%%%%%% m0.7tb5 %%%%%%%
\FIGURE[h]{
 \epsfig{file=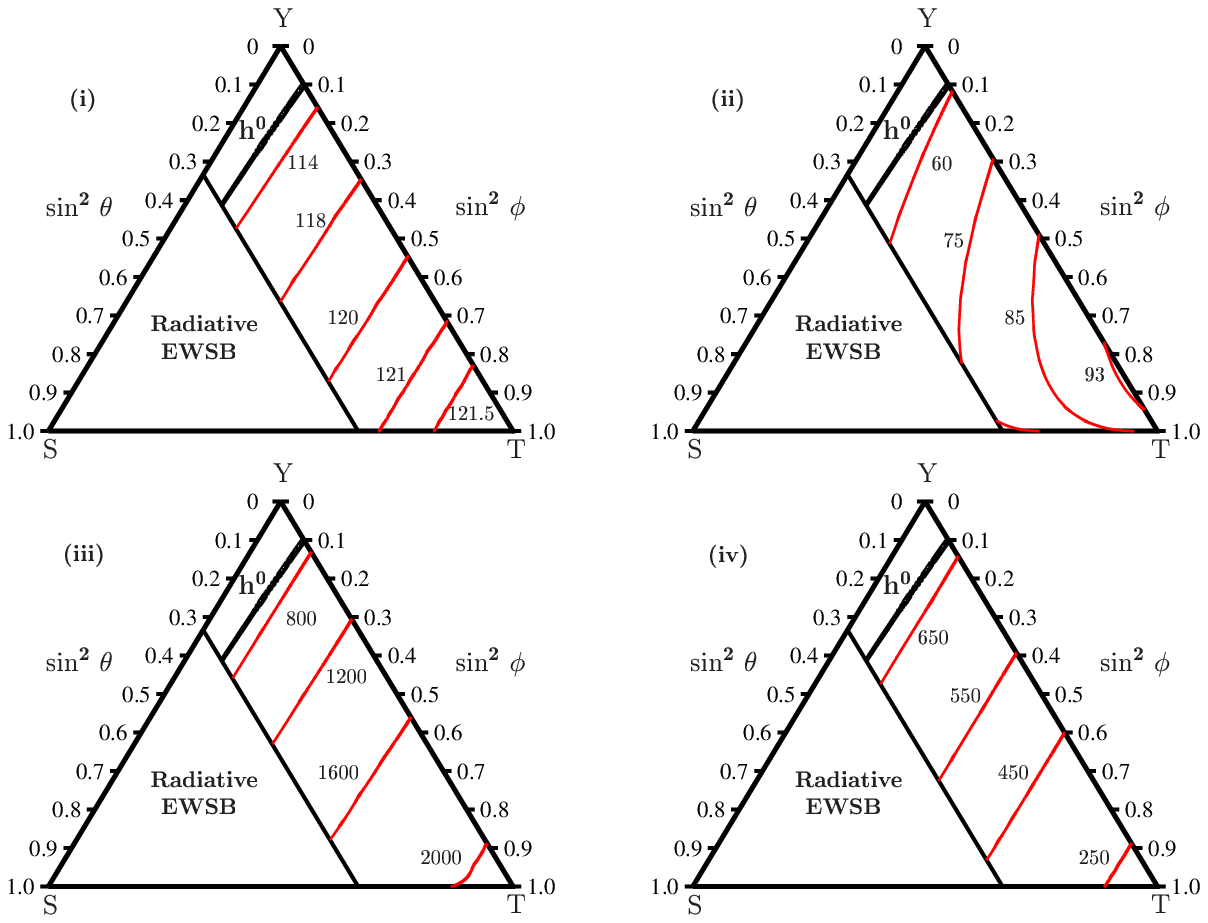,height=9.5cm}
 \caption{Sparticle mass contours (in GeV) for $m_{3/2}=700$
GeV and $\tan\beta=5$: (i) lightest higgs $h^0$, (ii)
neutralino $\chi_{1}^{0}$, (iii) gluino $\tilde{g}$ and (iv) stau
$\tilde{\tau}_1$.  The excluded regions are marked, and there is also
a tiny area close to untwisted moduli domination 
($\sin^2 \, \phi \approx 1.0$) where $\tilde{\tau}_1$ is the LSP.}
 \label{fig:trianglem0.7tb5}
}

Therefore for a
generic choice of $\theta$ and $\phi$ (away from any domination
limit), the relative contributions from each closed string F-term can
be easily determined by measuring the perpendicular distance of the
$\theta$--$\phi$ point away from each side of the triangle.  Note
that many earlier studies have not considered twisted moduli
contributions to SUSY breaking, and so these analyses are constrained
to the $S-T$ base of the triangle in Figure \ref{fig:tribasic}.

We have studied the allowed parameter spaces for various combinations
of $m_{3/2}$--$\tan\beta$ and sparticle mass contours are shown in
Figures \ref{fig:trianglem0.7tb5}-\ref{fig:anglem2tb20}.  We plot
the (i) lightest higgs $h^0$, (ii) neutralino $\chi_{1}^{0}$, (iii)
gluino $\tilde{g}$ and lightest squark/slepton state which is either
the (iv) stau $\tilde{\tau}_1$ or stop $\tilde{t}_1$.  
In Figure \ref{fig:trianglem0.7tb5} we observe that the viable
$\theta$--$\phi$ parameter space for $m_{3/2}=700$ GeV, 
$\tan\beta=5$ - which is not already forbidden by radiative symmetry breaking
- is only weakly constrained, although the region
with $\sin^2 \phi \simlt 0.3$ does not predict a sufficiently heavy Higgs
$h^0$.  Also there is a small region around the T-moduli
domination limit ($\sin^2 \theta \approx 0$, $\sin^2 \phi \approx 1$)
where the lightest stau state $\tilde{\tau}_1$ is the LSP 
and is ruled out by experimental constraints.  This is consistent with
the results shown in Figure \ref{fig:movedown} (a).
%\vskip-10mm
%%%%%% m1tb5 %%%%%%%
\FIGURE[h]{
 \epsfig{file=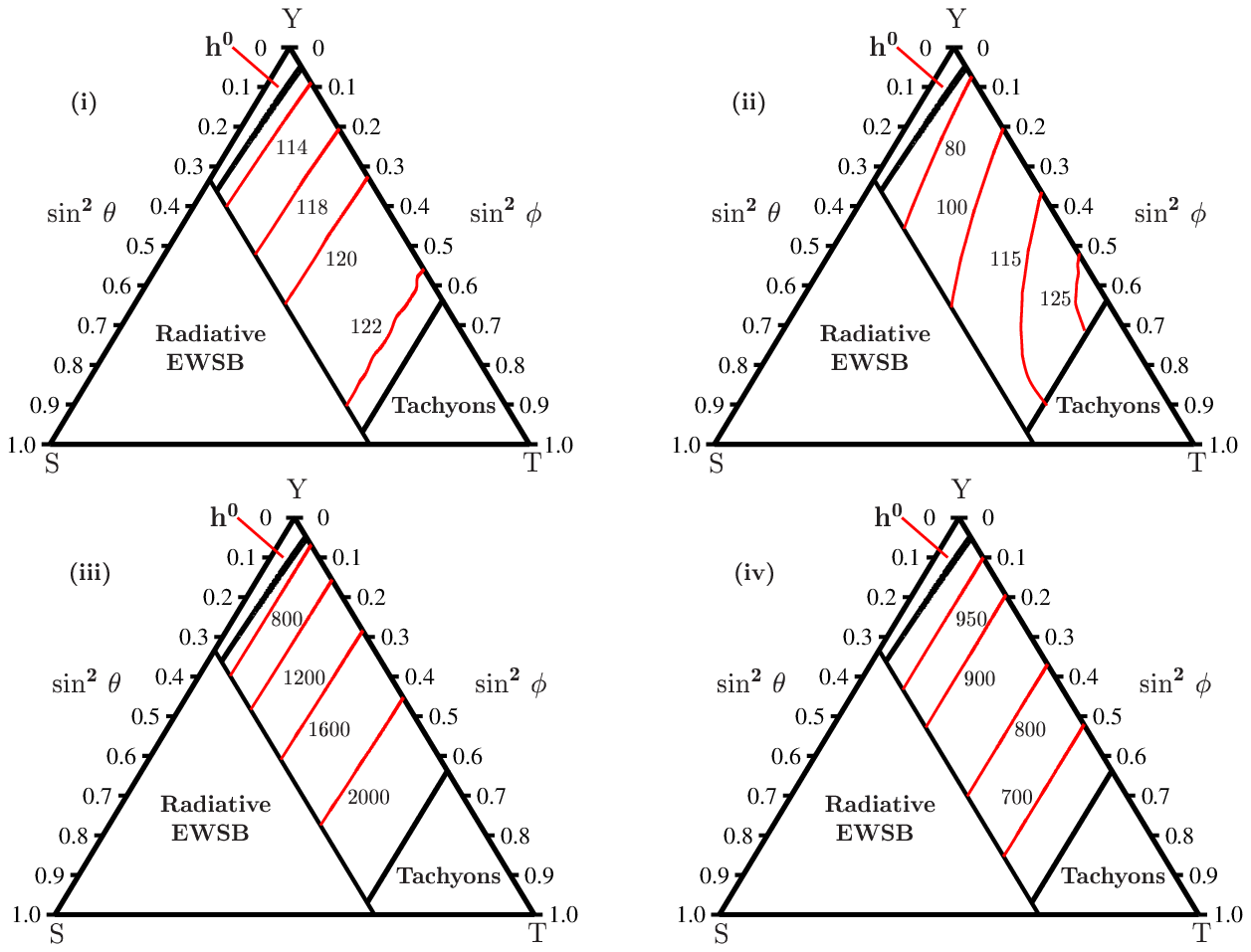,height=9.5cm}
 \caption{Sparticle mass contours (in GeV) for $m_{3/2}=1$ TeV
and $\tan\beta=5$: (i) lightest higgs $h^0$, (ii)
neutralino $\chi_{1}^{0}$, (iii) gluino $\tilde{g}$, (iv) stau
$\tilde{\tau}_1$.  The region with small $\phi$ is excluded by the
experimental Higgs bound, and large $\phi$ points are ruled out by
weak-scale tachyons.} 
 \label{fig:trianglem1tb5}
}

Comparing Figures \ref{fig:trianglem0.7tb5} and
\ref{fig:trianglem1tb5} demonstrates that even a slight increase in
$m_{3/2}$ can significantly reduce the parameter space by
truncating the viable region at larger values of $\phi$ due to the
presence of weak-scale tachyons.  Also the experimental Higgs
bound becomes less constraining since it is pushed to smaller $\phi$.
The range of viable Higgs and gluino masses is unaffected by the
increase in $m_{3/2}$, although the positions of the contours in
$\theta$--$\phi$ space are shifted.  However the neutralinos and
squark/slepton sparticles can be much heavier, and the lightest
squark/slepton state is typically the stau~\footnote{Notice that
the lightest stau $\tilde{\tau}_1$ contours are unique in that they
decrease in magnitude with increasing $\phi$.}.  There are also
regions with small $\sin^2 \phi \simlt 0.1-0.2$ in Figures 
\ref{fig:trianglem0.7tb5} and \ref{fig:trianglem1tb5} where the stop
squark mass is comparable to the stau.
%\vskip-10mm
%%%%%% m2tb20 %%%%%%%
\FIGURE[h]{
 \epsfig{file=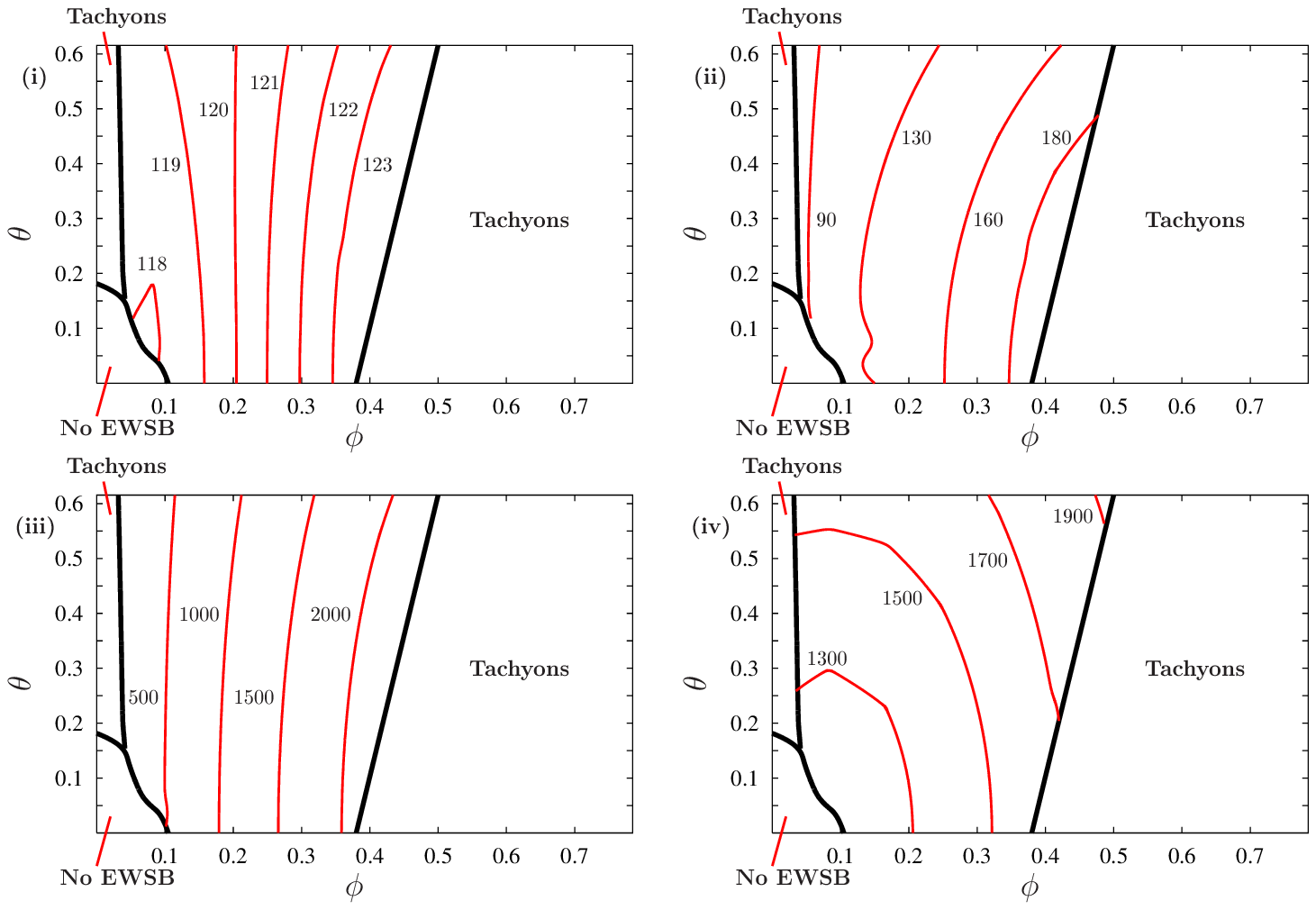,height=9.5cm}
 \caption{Sparticle mass contours (in GeV) for $m_{3/2}=2$ TeV
and $\tan\beta=20$: (i) lightest higgs $h^0$, (ii)
neutralino $\chi_{1}^{0}$, (iii) gluino $\tilde{g}$ and (iv) stop
$\tilde{t}_1$.  Note that we are using a reduced $\phi$ scale.}
 \label{fig:anglem2tb20}
}

When $m_{3/2}$ is raised further, we see that the viable
parameter space becomes increasingly narrow as weak-scale
tachyons sweep in from large $\phi$.  The small $\phi$ experimental
Higgs $h^0$ 
limit is replaced by another tachyon boundary, and a region opens up
around $\sin^2 \theta$, $\sin^2 \phi \approx 0$ where EWSB does
not occur. 
We also observe that the gross features of the viable parameter
space (for a given $m_{3/2}$) are not affected by increasing
$\tan\beta$, except that the no EWSB region grows with increasing
$\tan\beta$. 
Also note that for $m_{3/2}=1$ TeV, the effect of increasing
$\tan\beta$ replaces the Higgs $h^0$ bound in Figure
\ref{fig:trianglem1tb5} by a similar bound from the lightest chargino
$\chi_{1}^{0}$ and gluino $\tilde{g}$ sparticles. 
 
In Figure \ref{fig:anglem2tb20} we find that the viable parameter space 
for $m_{3/2}=2$ TeV, $\tan\beta=20$ is too small to represent within
the triangular framework, so we adopt the square
$\theta$--$\phi$ coordinates with a reduced $\phi$ axis.  The allowed
space is constrained by low-energy tachyons at both large and small
$\phi$, and also in the region around $\theta, \phi = 0$ where EWSB does not
occur.  The larger value of $\tan\beta$ allows the neutralino
$\chi_{1}^{0}$ and squark/slepton masses to be much heavier.  The stop
$\tilde{t}_1$ is now definitely lighter than the stau $\tilde{\tau}_1$
state, but it is still very heavy 
$m_{\tilde{t}_1} \sim {\cal O}( m_{3/2} )$.

%%%%%%%%%%%%%%%%%%%%%%%%%%%%%%%%%%%%%%%%%%%
\subsection{Benchmarks}  \label{sec:benchmarks}

Recently there have been many studies of ``benchmark'' 
points that highlight particular characteristics within a model
parameter space~\footnote{See Refs.~\cite{snowmass,kane} for recent examples.}.
In Table \ref{tab:benchmark} we list some sparticle spectra for our
own set of benchmark points.
Point A is the twisted moduli domination limit of section
\ref{sec:t0p0}, and we also list the more general points B and C from 
section \ref{sec:general} where the dilaton and moduli F-terms each
contribute to SUSY breaking.  The full outputs for these three sample
points can be found on 
\begin{quote}
 {\tt http://allanach.home.cern.ch/allanach/benchmarks/int.html}
\end{quote} 
in the Les Houches Accord output format~\cite{leshouches} to enable
future Monte-Carlo analyses.

%\vskip-3mm
%\newpage
\TABLE{
 \scalebox{0.9}{
\begin{tabular}{|c||c|c|c|} \hline
   Point & A & B & C \\ \hline
   $\theta$ & 0 & 0.1 & 0.615 \\
   $\phi$ & 0 & 0.1 & 0.1 \\
   $m_{3/2}$ & 5000 & 2000 & 2000 \\ 
   $\tan\beta$ & 4 & 10 & 20 \\ \hline\hline
% HIGGS
   $m_{h^0}$     & 114.7 & 117.1 & 118.9 \\
   $m_{A^0}$     & 5233  & 1991  & 1500 \\
   $m_{H^0}$     & 5233  & 1996  & 1528 \\ 
   $m_{H^{\pm}}$ & 5234  & 1992  & 1513 \\ \hline
% NEUTRALINOS + CHARGINOS + GLUINO
   $m_{\tilde{\chi}_1^0}$     & 98.5 & 123 & 104 \\
   $m_{\tilde{\chi}_2^0}$     & 273  & 179 & 155 \\
   $m_{\tilde{\chi}_1^{\pm}}$ & 98.5 & 178 & 155 \\
   $m_{\tilde{g}}$            & 586  & 509 & 428 \\ \hline
% STOP + SCHARM + SUP 
   $m_{\tilde{t}_1}$ & 3061 & 1196 & 1549 \\
   $m_{\tilde{t}_2}$ & 4182 & 1668 & 1789 \\
   $m_{\tilde{c}_1}$, $m_{\tilde{u}_1}$ & 470 & 2314 & 1716 \\
   $m_{\tilde{c}_2}$, $m_{\tilde{u}_2}$ & 482 & 2318 & 1719 \\
    \hline
% SBOTTOM + SSTRANGE + SDOWN 
   $m_{\tilde{b}_1}$ & 4189 & 1666 & 1780 \\
   $m_{\tilde{b}_2}$ & 5019 & 2002 & 1977 \\
   $m_{\tilde{s}_1}$, $m_{\tilde{d}_1}$ & 476 & 2314 & 1716 \\
   $m_{\tilde{s}_2}$, $m_{\tilde{d}_2}$ & 478 & 2319 & 1721 \\ \hline
% STAU + SMU + SELECTRON
   $m_{\tilde{\tau}_1}$ & 4999 & 1978 & 1947 \\
   $m_{\tilde{\tau}_2}$ & 5000 & 1989 & 1983 \\
   $m_{\tilde{\mu}_1}$, $m_{\tilde{e}_1}$  & 150 & 2273 & 1681 \\
   $m_{\tilde{\mu}_2}$, $m_{\tilde{e}_2}$  & 239 & 2276 & 1684 \\ 
    \hline 
% SNEUTRINO
   $m_{\tilde{\nu}_e}$, $m_{\tilde{\nu}_{\mu}}$ & 131 & 2275 & 1682 \\
   $m_{\tilde{\nu}_{\tau}}$ & 4999 & 1988 & 1975 \\ \hline
% LSP
   LSP & $\tilde{\chi}_1^0$ & $\tilde{\chi}_1^0$ & $\tilde{\chi}_1^0$ \\ 
    \hline     
\end{tabular}}
\caption{Sample spectra: All masses are in GeV and
 the Goldstino angles are given in radians.  The high-scale input soft
 parameter can be found in Tables \ref{tab:moveup},\ref{tab:t0_1p0_1}
 and \ref{tab:nohiggs} by multiplying by $m_{3/2}$.}
\label{tab:benchmark}
}

%%%%%%%%%%%%%%%%%%%%%%%%%%%%%%%%%%%%%%%%%%%%%%%%%%%%%%%%%%
%%%%%%%%%%%%%%%%%%%%%%%%%%%%%%%%%%%%%%%%%%%%%%%%%%%%%%%%%%
%%%%%%%%%%%%%%% SECTION: CONCLUSIONS %%%%%%%%%%%%%%%%%%%%%
%%%%%%%%%%%%%%%%%%%%%%%%%%%%%%%%%%%%%%%%%%%%%%%%%%%%%%%%%%
\section{Discussion and Conclusions}  \label{sec:disc}

The original parallel-brane models of sequestering~\cite{amsb,gmsb} have many
appealing features for supersymmetric phenomenology.  They offer an
attractive and simple realisation of traditional hidden-sector models
by localising the Standard Model and SUSY breaking sectors at opposite
ends of a higher-dimensional bulk spacetime.  A sequestered \kahler
potential, reminiscent of no-scale SUGRA models~\cite{noscale},
predicts that 
there are no direct couplings between the two sectors at leading
order.  Non-renormalisable higher-dimensional operators can be
constructed that are found to give only exponentially small contributions
using EFT techniques.  Therefore the dominant
contributions to the soft masses and trilinears arise from
radiative corrections involving bulk fields.  The absence of
significant couplings may alleviate the supersymmetric flavour problem
since small off-diagonal squark/slepton mass-matrix elements arise at
the weak-scale through calculable RGE running effects.
Unfortunately the simplest models of sequestering have severely
constrained parameter spaces, and there have been recent concerns that
the sequestered \kahler potential is unnatural in realistic string
constructions~\cite{dine}.  

This motivated us to propose a string-inspired gaugino mediation model of 
sequestering with twisted moduli~\cite{benakli,tmsb} 
to tackle these problems.  We have attempted to put sequestering on a firmer
theoretical basis by embedding it in a type I 
construction involving intersecting D-branes, where localised closed
string twisted moduli play the r\^{o}le of the SUSY breaking sector
that can be spatially-separated away from open string matter
fields.  We proposed a modified \kahler potential for the intersection
states representing MSSM fields~\cite{tmsb} that attributes the sequestering
suppression factor (in the limit of twisted moduli domination) to
non-perturbative instanton effects involving strings stretching
between different fixed points~\cite{altexp}.  As we have discussed in
this paper, there are
additional sources of SUSY breaking from the dilaton and untwisted
T-moduli states that can be interpreted as gravity mediated SUSY
breaking, and we have used a Goldstino angle~\cite{ibanez94}-\cite{ibanez98}
parameterisation to control the relative contributions to the overall
SUSY breaking.  This parameterisation has allowed us to 
move away smoothly from the sequestered limit -- where only the twisted
moduli contribute to SUSY breaking -- and study the interplay between
sequestering and gravity mediation by including the dilaton/T-moduli
contributions.  
%In this way we observe that the severely constrained
%parameter space for twisted moduli domination can be opened up with the
%additional gravity mediation effects.

In this paper we have studied the twisted moduli domination limit,
where we have observed that the  
Higgs mass is light due to restricted $\tan\beta$ from the EWSB
constraint, and where the large third family soft scalar masses yield a
characteristic ``stop-heavy MSSM'' sparticle spectrum.  In this case
we have found a wino-dominated LSP which results in quasi-degenerate
lightest charginos and neutralinos, leading to an opportunity, as well
as difficulties in the detection of charginos~\cite{barr}.  We have
also considered additional contributions from the untwisted T-moduli
to the twisted moduli limit, and seen that the bounds on $\tan\beta$
become relaxed allowing for a larger region of viable parameter space,
where the extra contributions destroy the quasi-degeneracy between
chargino and neutralino. We have seen that the converse is also  
true, namely that the twisted moduli domination limit is strongly
constrained by the tachyons, but adding T-moduli contributions
considerably opens up the parameter space.  
We also study a particular mixed point with contributions from both
the dilaton and moduli sectors that predicts a
vanishing Higgs doublet soft mass at the GUT-scale without coming into
conflict with empirical bounds, and we find that the lightest stop can
become very heavy.

The experimental constraints from FCNC
processes are satisfied within our model by two different mechanisms.
Firstly in the twisted moduli domination limit, it is 
important to emphasise that the first two squark and slepton 
families have approximately zero soft masses at the high energy
scale due to sequestering.  This leads to a natural suppression of
the most dangerous 
flavour changing neutral currents involving the first two families.
The reason is that, at low energies, diagonal soft masses are generated 
for the first two families via RGE effects
proportional to gauge couplings which are flavour blind, 
resulting in a universal form for the low energy soft mass matrices for the
first and second family squarks and sleptons proportional
to the unit matrix. This mechanism is very similar to that which 
operates in the original \gmsb scenario, however
unlike that scenario, where all three families of squarks and
sleptons have approximately zero soft masses at the high energy scale,
here this applies to only the first two families.  Nevertheless this
proves sufficient to suppress the most dangerous FCNCs associated
with the first two families as our results in section \ref{sec:fcnc}
demonstrate. 
 
The second mechanism suppresses FCNCs when we include
gravity mediation effects to move away from twisted moduli domination.
We observe that the viable parameter space rapidly increases and
yields heavy sparticle masses.  However the additional contributions
to SUSY breaking also spoil the sequestering such that first and
second family soft scalar masses are no 
longer negligible at the high-scale, and the FCNCs cannot be
suppressed as before.  Instead the heavier gluino and squark masses
ameliorate the SUSY flavour problem with the additional
assumption of family-diagonal sfermion mass matrices (in the weak-basis) at the
high-scale.  
In Table \ref{tab:mid}, we see that some of the MI deltas
$(\delta^{u}_{12})_{LL}$ and 
$\sqrt{(\delta^{u}_{12})_{LL} (\delta^{u}_{12})_{RR}}$ 
are within a factor of a few to 100 of being probed by experiment when
we assume that the weak-scale Yukawa mixing occurs exclusively in the
up-quark sector.  If we change this assumption and move the mixing
into the down-quark sector, we find that the previous $\delta^u_{12}$
values are reduced while the equivalent $\delta^d_{12}$  MI
deltas become larger.  In fact this new value for 
$\sqrt{(\delta^{d}_{12})_{LL} (\delta^{d}_{12})_{RR}}$ effectively
rules out the twisted moduli domination limit, albeit with weak-scale
mixing exclusively in the down-quark sector.

To summarise, we have performed a first phenomenological
analysis of the phenomenology of twisted
moduli in type I string theory. Within this framework we
have discussed a string inspired version of gaugino
mediated supersymmetry breaking where only the
third squark and slepton family feels directly the supersymmetry breaking
effects of the twisted moduli, and is therefore predicted to
be significantly heavier than the first two squark and slepton
families which are massless at high energies in the
twisted moduli domination limit. More generally we have
considered the smooth interpolation between such sequestered
scenarios and gravity-mediated scenarios, by switching
on the supersymmetry breaking 
effects of non-twisted moduli such as $S$ and $T_i$ in addition
to the twisted moduli $Y_2$, which opens up the parameter
space considerably. If SUSY is discovered at the 
LHC it will be interesting to see if the spectrum of
superpartners corresponds to any of the general
type I spectra including the new 
sequestered effects of twisted moduli considered in this paper.

%In this work we have made the first steps towards combining
%sequestering with gravity mediated SUSY breaking within a type I
%string framework.  We have proposed a phenomenologically-motivated
%\kahler potential for intersection states and studied the resulting
%phenomenology. 

%%%%%%%%%%%%%%%%%%%%%%%%%%%%%%%%%%%%%%%%%%%%%%%%%%%%%%%%%%
%%%%%%%%%%%%%%% ACKNOWLEDGEMENTS %%%%%%%%%%%%%%%%%%%%%%%%%
%%%%%%%%%%%%%%%%%%%%%%%%%%%%%%%%%%%%%%%%%%%%%%%%%%%%%%%%%%
\acknowledgments
This work is partially supported by PPARC.  BCA would like to thank
Karim Benakli for an enlightening discussion.  The work of D.R. is
supported by the RTN European Program HPRN-CT-2000-00148.  D.R. would
like to thank Anna Rossi and Sudhir Vempati for useful discussions.

%%%%%%%%%%%%%%%%%%%%%%%%%%%%%%%%%%%%%%%
\section*{Note Added}

During the final stages of this paper, a phenomenological study~\cite{baer}
appeared on the archive which considers lighter SUSY breaking masses for
the first two families. There, it is shown how the lighter two families
can reconcile $(g-2)_\mu$, dark matter and $b \longrightarrow s \gamma$
constraints compared to the usual universal minimal SUGRA assumption. There is
little overlap with the present paper except for the heavier third
family sfermions, but we note that their conclusions
should qualitatively apply to our model also, adding support to it.

%%%%%%%%%%%%%%%%%%%%%%%%%%%%%%%%%%%%%%%%%%%%%%%%%%%%%%%%%%
%%%%%%%%%%%%%%%%%%%%%%%%%%%%%%%%%%%%%%%%%%%%%%%%%%%%%%%%%%
%%%%%%%%%%%%%%% SECTION: APPENDIX %%%%%%%%%%%%%%%%%%%%%%%%
%%%%%%%%%%%%%%%%%%%%%%%%%%%%%%%%%%%%%%%%%%%%%%%%%%%%%%%%%%
\appendix
\section{Derivation of the SUSY breaking F-terms}   \label{app:softterms}

In this appendix we will show how the SUSY breaking F-terms are
derived in terms of Goldstino angles using a series expansion in
inverse powers of $T_2+\overline{T}_2$.  We follow the analysis in 
Refs.~\cite{ben,ben2}, but generalised to three untwisted moduli $T_i$
and $D5_i$-branes.  Recall that the K\"{a}hler potential for the
dilaton and moduli (after suppressing $M_{\ast}$) is:  
\begin{eqnarray}
 K(S, \overline{S}, T_i, \overline{T_i}, Y_2, \overline{Y_2})
  = - \ln (S + \overline{S}) - \sum_{i=1}^{3} \ln (T_i + \overline{T_i}) 
   + \hat{K} 
     \label{eq:kp2}
\end{eqnarray}  
where we choose to leave the precise form of the twisted moduli K\"{a}hler
potential as an unknown function $\hat{K} \equiv \hat{K}(X)$ with an argument
$X = Y_2 + \overline{Y_2} - \delta_{GS} \, {\mathrm ln} ( T_2 +
\overline{T_2} )$ required to cancel modular anomalies~\cite{gs}.  The
mixing induced between $T_2$ and $Y_2$ by this unknown K\"{a}hler potential
means that the K\"{a}hler metric is no longer diagonal (if we ignore
open string matter fields).  In order to invert the K\"{a}hler
metric $K_{\overline{J} \, I}$ (where $I, J = S, T_i, Y_2$ label
derivatives), we invoke the unitarity relation 
\begin{eqnarray}
 P^{\dagger} \,K_{\overline{J} \, I} \, P = 1
  \label{eq:punitary}
\end{eqnarray}
where the matrix $P$ canonically normalises $K_{\overline{J} \, I}$. 

\noindent From Eq.(\ref{eq:kp2}), we obtain the following non-zero
entries in the K\"{a}hler metric:
\begin{eqnarray}
 K_{\overline{S} \, S} = \frac{1}{(S+\overline{S})^2} 
  \hspace*{6mm}, \hspace*{6mm}  
 K_{\overline{T_i} \, T_i} = \frac{1}{(T_i+\overline{T_i})^2}
  \hspace*{4mm} (i=1,3) 
   \hspace*{6mm}, \hspace*{6mmm} 
 K_{\overline{Y_2} \, Y_2} = \hat{K}''  \hspace*{6mm} \label{eq:km}\\
 K_{\overline{T_2} \, T_2} = \frac{1}{(T_2+\overline{T_2})^2} \left[
  1 + \delta_{GS} \hat{K}' + \delta_{GS}^{2} \hat{K}'' \right]
   \hspace*{6mm}, \hspace*{6mm} 
 K_{\overline{Y_2} \, T_2} = K_{\overline{T_2} \, Y_2} 
  = - \frac{\delta_{GS}}{T_2+\overline{T_2}} \, \hat{K}''  \nonumber
\end{eqnarray}
where $\hat{K}'$ ($\hat{K}''$) means the first (second) derivative
with respect to the argument $X = Y_2 + \overline{Y_2} - \delta_{GS} \, 
{\mathrm ln} ( T_2 + \overline{T_2} )$.  

\noindent The full K\"{a}hler metric is:
\begin{eqnarray}
  K_{\bar{J} I} = \left(
  \begin{array}{ccccc}
   \frac{1}{(S+\overline{S})^{2}} & 0 & 0 & 0 & 0 \\
   0 & \frac{1}{(T_{1}+\overline{T_1})^{2}} & 0 & 0 & 0 \\
   0 & 0 & \frac{1}{(T_{2}+\overline{T_2})^{2}} \left( k +
    \delta_{GS}^{2} \right) & 0 & 
     - \frac{\delta_{GS}}{T_{2}+\overline{T_2}} \\
   0 & 0 & 0 & \frac{1}{(T_{3}+\overline{T_3})^{2}} & 0 \\
   0 & 0 & - \frac{\delta_{GS}}{T_{2}+\overline{T_2}} & 0 & \hat{K}''
  \end{array} \right)  \label{eq:kmat}
\end{eqnarray}
where $I=S,T_{i},Y_{2}$ and for convenience we have introduced the unphysical
parameter $k$ which is given by:
\begin{eqnarray}
 k = (T_{2}+\overline{T_2})^{2}  K_{ \overline{T_2} \, T_2} - \delta_{GS}^2
 = 1 + \delta_{GS} \, \hat{K}'  \label{eq:kpar}
\end{eqnarray}
where we have made a simplifying assumption and set $\hat{K}'' = 1$ 
in agreement with Refs.~\cite{ben,ben2}.

Using the standard relation between F-terms, gravitino mass and the
matrix $P$ \cite{ibanez98}, we define the F-terms using Goldstino angles as: 
\begin{eqnarray}
 F_{I} \equiv \left(
  \begin{array}{c}
   F_{S} \\ F_{T_{1}} \\ F_{T_{2}} \\ F_{T_{3}} \\ F_{Y_{2}} 
  \end{array} \right)
 = \sqrt{3} \, m_{3/2}  \, P \left(
  \begin{array}{c}
   \sin \theta \\
   \cos \theta \, \sin \phi \, \Theta_{1} \\
   \cos \theta \, \sin \phi \, \Theta_{2} \\
   \cos \theta \, \sin \phi \, \Theta_{3} \\
   \cos \theta \, \cos \phi \\
  \end{array} \right)
   \label{eq:Fterms}
\end{eqnarray}
where $\theta$ and $\phi$ are Goldstino angles, 
$\sum_{i} \Theta_{i}^{\dagger} \Theta_{i} = 1$ and we have not included
any CP-phases.

Using Eqs.(\ref{eq:punitary}) and (\ref{eq:kmat}), we obtain an
expression for the matrix $P$ that can be expanded for large values of
($T_{2}+\overline{T}_{2}$) to give:
\begin{eqnarray}
  P = \left(
  \begin{array}{ccccc}
   S+\overline{S} & 0 & 0 & 0 & 0 \\
   0 & T_{1}+\overline{T}_{1} & 0 & 0 & 0 \\
   0 & 0 & \frac{T_{2}+\overline{T}_{2}}{\sqrt{k}} & 0 & 
    - \frac{\delta_{GS}}{T_{2}+\overline{T}_{2}} \\
   0 & 0 & 0 & T_{3}+\overline{T}_{3} & 0 \\
   0 & 0 & \frac{\delta_{GS}}{\sqrt{k}} + \frac{\sqrt{k} 
    \delta_{GS}}{(T_{2}+\overline{T}_{2})^{2}} & 0 & 1 
     - \frac{\delta_{GS}^{2}}{(T_{2}+\overline{T}_{2})^{2}}
  \end{array} \right)
 + {\mathcal O} \left[ \frac{1}{(T_{2}+\overline{T}_{2})^{3}} \right]
  \hspace*{5mm} \label{Pmatrix}
\end{eqnarray}
which leads to the following SUSY breaking F-terms from Eq.(\ref{eq:Fterms}):
\begin{eqnarray}
 F_{S} &=& \sqrt{3} m_{3/2} \sin \theta \, (S+\overline{S}) \nonumber \\
 F_{T_{1}} &=& \sqrt{3} m_{3/2} \cos \theta \sin \phi \, \Theta_{1} 
  \, (T_{1} + \overline{T}_{1}) \nonumber \\
 F_{T_{2}} &=& \sqrt{3} m_{3/2} \cos \theta \left[ \sin \phi \frac{(T_{2}+
  \overline{T}_{2})}{\sqrt{k}} \, \Theta_{2} 
   - \cos \phi \, \frac{\delta_{GS}}{T_{2} +\overline{T}_{2}} 
    \right]  \label{eq:fterms3} \\
 F_{T_{3}} &=& \sqrt{3} m_{3/2} \cos \theta \sin \phi \, \Theta_{3} \,
  (T_{3} + \overline{T}_{3}) \nonumber \\
 F_{Y_{2}} &=& \sqrt{3} m_{3/2} \cos \theta \left[ \sin \phi \left( 
  \frac{\delta_{GS}}{\sqrt{k}} + \frac{\sqrt{k} \delta_{GS}}{(T_{2}
   +\overline{T}_{2})^{2}} \right) \Theta_{2} 
    + \cos \phi \left( 1 
     - \frac{\delta_{GS}^{2}}{(T_{2} +\overline{T}_{2})^{2}} \right) \right] 
  \nonumber 
\end{eqnarray}
where $F_{T_{2}}$ and $F_{Y_{2}}$ are expanded up to 
${\mathcal O} \left[ \frac{1}{(T_{2}+\overline{T}_{2})^{3}} \right]$
and $k = 1 + \delta_{GS} \, \hat{K}'$.  Notice that setting
$\delta_{GS}=0$ removes the mixing between $T_2$ and $Y_2$ to recover
the standard F-term expressions of Ref.~\cite{ibanez98}.

%%%%%%%%%%%%%%%%%%%%%%%%%%%%%%%%%%%%%%%%%%%%%%%%%%%%%%%%%%
%%%%%%%%%%%%%%%%%%%%%%%%%%%%%%%%%%%%%%%%%%%%%%%%%%%%%%%%%%
\section{Phenomenological problems with ``pure'' \gmsb}  \label{app:gmsb}

In this appendix we will consider the original \gmsb model within our
D-brane setup involving
three degenerate families sequestered away from SUSY breaking along an
extra dimension~\cite{gmsb}.  It is well known that the \gmsb
scenarios are severely constrained, and we would like to see if the
additional contributions from the dilaton and untwisted moduli
(gravity mediation) offer a solution.  Instead we observe that
sequestering the third family leads to severe problems that
effectively rule out ``pure'' \gmsb within our SI$\tilde{g}$M
framework. 
%\vskip-7mm
\FIGURE[h]{
 \epsfig{file=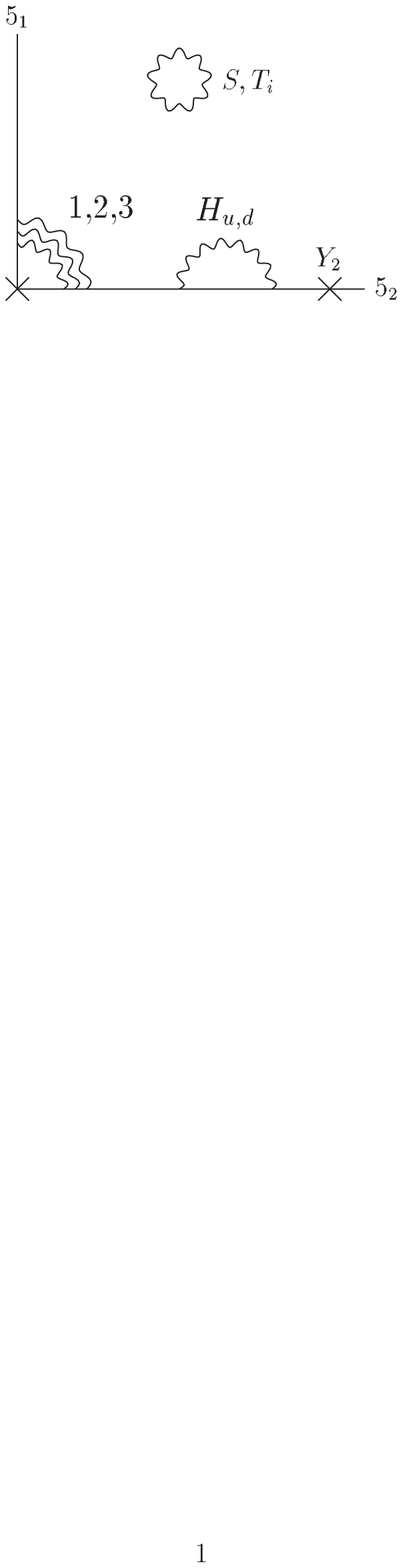,height=4.2cm}
 \caption{The allocation of open and closed string states in a
string-inspired \gmsb model of the MSSM.  The three degenerate MSSM
families ($1,2,3$) are localised at the intersection $\ci$, and the
Higgs doublets $H_{u,d}$ are identified as $\ctf$ states.}
 \label{fig:modelgmsb}
}
Our starting point is a string-inspired model with all three MSSM
families localised at the intersection region between D5-branes as
shown in Figure \ref{fig:modelgmsb}, where the MSSM gauge group is once
more dominated by the components on the D$5_2$-brane in the
single-brane dominance limit.  In order to have at least a top
Yukawa coupling at (renormalisable) leading order, we choose the
following assignment of MSSM states:
\begin{eqnarray}
 Q_{iL} \, , \, L_{iL} \, , \, U_{iR}^c \, , \, D_{iR}^c 
  \, , \, E_{iR}^c &\equiv& C^{5_1 5_2} \hspace*{1cm} (i=1,2,3) 
    \label{eq:gmsball}  \\
 H_{u} \, , \, H_d &\equiv& C^{5_2}_3  \nonumber
\end{eqnarray}
which yields a democratic Yukawa texture from the allowed
superpotential term
\begin{eqnarray}
 W_{ren} = \ctf \, \ci \, \ci   \label{eq:wren2}
\end{eqnarray}
In order to compare with the earlier analysis with only two
sequestered families, we will assume that
some (unspecified) flavour symmetry allows us to rotate the democratic
texture into hierarchical form with a single universal non-zero (33)
entry.  

\vskip5mm
\noindent
$\bullet$ {\bf Soft Parameters:}
\vskip2mm

We follow the same arguments as section \ref{sec:sbp} to
calculate the soft parameters that provide GUT-scale boundary
conditions for the RGE analysis.  We will assume that the universal squark and
slepton mass-squared matrices are diagonal in family space at the high-scale:
\begin{eqnarray}
 m_{\tilde{Q}}^{2} \, , \, m_{\tilde{L}}^{2} \, , \, m_{\tilde{u}}^{2}
  \, , \, m_{\tilde{d}}^{2} \, , \, m_{\tilde{e}}^{2} = \left( 
   \begin{array}{ccc} 
    m_{\ci}^2 & 0 & 0 \\
    0 & m_{\ci}^2 & 0 \\
    0 & 0 & m_{\ci}^2 
   \end{array}  \right)  \label{eq:ssmsgmsb}
\end{eqnarray}
where the universal scalar mass-squared $m_{\ci}^2$ is given by
Eq.(\ref{eq:softci}).  The Higgs doublets soft scalar mass-squared
are also universal at the high-scale:
\begin{eqnarray}
 m_{H_u}^2 \, , \, m_{H_d}^2 \equiv
  m_{C^{5_2}_3}^2 = m_{3/2}^2 \left( 1 - \cos^2 \theta \, \sin^2 \phi \right)
    \label{eq:softc23}
\end{eqnarray}
The soft gaugino masses remain unchanged from Eq.(\ref{eq:softga}).
However the soft trilinear takes a more complicated form, and we still
impose a (33) hierarchical texture:
\begin{eqnarray}
 A^f_{ij} \, Y^f_{ij}  = \left(
  \begin{array}{ccc}
   0 & 0 & 0 \\
   0 & 0 & 0 \\
   0 & 0 & \hat{{\cal A}} \, Y^f_{33}
  \end{array} \right)  \hspace*{1cm} (f = u, d, e)
    \label{eq:tribcgmsb}
\end{eqnarray}
where $Y^f_{33}$ is the (33) element of the (running) Yukawa matrix
$Y^f$ at the GUT-scale and the universal soft trilinear $\hat{{\cal
A}}$ is: 
\begin{eqnarray}
 \hat{{\cal A}} \equiv 
  A_{\ctf \ci \ci} = - m_{3/2} \cos\theta \left[ \sin\phi \left( 1 
   + \frac{X^2}{12} \, e^{-(T_{2}+\bar{T}_{2})/4} \,
    (T_{2}+\bar{T}_{2}) \right)  \right.  \hspace*{2cm}
      \label{eq:softtrigmsb}  \\
 + \left. \frac{2 X}{\sqrt{3}} \, \cos\phi \,
  \left( 1 - e^{-(T_{2}+\bar{T}_{2})/4} \right) \right] 
 + {\cal O} \left[ \frac{\delta_{GS} \, X \,
  e^{-(T_{2}+\bar{T}_{2})/4} }{(T_2 + \overline{T_2})} \right]  \nonumber
\end{eqnarray}
where $X = Y_2 + \overline{Y_2} 
 - \delta_{GS} \, {\mathrm ln} ( T_2 + \overline{T_2} )$.

\vskip5mm
\noindent
$\bullet$ {\bf Phenomenological problems with a large trilinear:}
\vskip2mm

Following the gauge coupling unification arguments in section
\ref{sec:choose} and the choice of parameters in Table
\ref{tab:paramcon}, we find that the soft trilinear is numerically: 
\begin{eqnarray}
 \hat{{\cal A}} 
  \approx - m_{3/2} \cos\theta \left[ \sin\phi + 45 \cos\phi \right] 
    \label{eq:softtrigmsblimit}
\end{eqnarray}
where the higher-order terms in Eq.(\ref{eq:softtrigmsb}) can be
safely neglected.  Notice that this soft trilinear has an additional (large)
piece $\sim (\cos\theta \cos\phi)$ in contrast to the trilinear in
Eq.(\ref{eq:softtri}), and we will show that this extra piece leads to problems
with sfermion tachyons.

For example, we will focus on the stop sector in the limit of twisted
moduli domination ($\theta, \, \phi = 0$) where the trilinear is non-zero
$\hat{{\cal A}} \approx - 45 \, m_{3/2}$.  Assuming negligible
intergenerational scalar mixing, the tree-level stop mass-squared eigenvalues
$m_{\tilde{t}_{1,2}}^2$ are found by diagonalising the following 
$2 \times 2$ mass-squared matrix ${\cal M}_{\tilde{t}}^2$ in the 
$(\tilde{t}_L \,\, \tilde{t}_R)^T$ basis:
\begin{eqnarray}
 {\cal M}_{\tilde{t}}^2 = 
  \left( 
   \begin{array}{cc}
    \left( m_{\tilde{Q}}^2 \right)_{33} + m_t^2 + \ldots &
    m_t \left( \hat{{\cal A}} - \mu \, \cot\beta \right) \\
    \vrule height 18pt width 0pt  
    m_t \left( \hat{{\cal A}} - \mu \, \cot\beta \right) &
    \left( m_{\tilde{u}}^2 \right)_{33} + m_t^2 + \ldots 
   \end{array}
  \right)  
   \label{eq:stopm2}
\end{eqnarray}
where we have dropped $M_{Z^0}^2$ terms; $m_t$ is the physical top
quark mass; and $\left( m_{\tilde{Q}}^2 \right)_{33} \, , \,
\left( m_{\tilde{u}}^2 \right)_{33} = m_{\ci}^2$ from
Eq.(\ref{eq:ssmsgmsb}).  The determinant of Eq.(\ref{eq:stopm2}) is
the product of the mass-squared eigenvalues: 
\begin{eqnarray}
 m_{\tilde{t}_{1}}^2 \, m_{\tilde{t}_{2}}^2 =  
  \left( m_{\ci}^2 + m_t^2 \right)^2 
   - m_t^2 \left( \mu - \hat{{\cal A}} \, \cot\beta \right)^2 + \ldots
     \label{eq:stopm2det}
\end{eqnarray}
which will be negative if $\hat{{\cal A}}$ is too large (as happens
here).  This implies that the scalar potential involves a tachyon mass
term that rules out the parameter space.

In order to avoid these tachyon problems, we must reduce the magnitude
of $\hat{{\cal A}}$, and in particular the factor:
\begin{eqnarray}
 \frac{2}{\sqrt{3}} \,  
  \left( 1 - e^{-(T_{2}+\bar{T}_{2})/4} \right)
   \left[ Y_2 + \overline{Y_2} 
    - \delta_{GS} \, {\mathrm ln} ( T_2 + \overline{T_2} ) \right] \approx 45
  \label{eq:factor}
\end{eqnarray}
when we input the parameter choices from Table \ref{tab:paramcon} into
Eq.(\ref{eq:softtrigmsb}).  Our choice of $(Y_2+\overline{Y}_2) = 0$
and $(T_2+\overline{T}_2) = 50$ are motivated by gauge coupling
unification in the single-brane dominance limit; the
reliability of the series expansion of F-terms in
Eq.(\ref{eq:fterms3}); and we also need a sufficiently large value of 
$(T_2+\overline{T}_2)$ for sequestering.  
There is some flexibility in the choice of $\delta_{GS}$, and
reducing its value will indeed lower $\hat{{\cal A}}$.  However the
effect is not enough to make the matrix-determinant in Eq.(\ref{eq:stopm2det})
positive since all three families of sfermions are sequestered with
exponentially-suppressed masses.  

In contrast, the original parallel-brane \gmsb models~\cite{gmsb} do
have viable regions of parameter space with sequestered squark/slepton
masses, vanishing trilinears and gauginos/Higgses soft masses $\sim
{\cal O}(m_{3/2})$.  However in
our framework the gauginos are too light to help, although they can be made
heavier by increasing $\delta_{GS}$, but we already know that this
will aggravate the tachyon problem further by increasing the trilinear
$\hat{{\cal A}}$.  

We must conclude that ``pure'' \gmsb (in the twisted moduli domination
limit) is not viable within the D-brane setup shown in Figure
\ref{fig:modelgmsb} and the choice of model parameters given in Table
\ref{tab:paramcon}.  An alternative solution may be to introduce
contributions from the dilaton and untwisted moduli sectors, but we
find that this requires large Goldstino angles $\theta, \phi \simgt
0.5$.  We could also consider different allocations of open string
states, but this also leads to problems.  For instance, we could
choose to sequester the Higgs doublets on the D$5_1$-brane with the
three MSSM families kept as intersection states $\ci$.  The dominant
top Yukawa coupling can arise from a term in the renormalisable
superpotential $W_{ren} \sim \cof \, \ci \, \ci$ which automatically
forbids R-parity violating superpotential terms at leading order.
Unfortunately the asymmetric compactification with $\Rt \gg \Ro$ -- that
allows gauge coupling unification on the D$5_2$-brane -- implies 
that $\gt^2 \ll \go^2$, and so the Higgs doublets will have a different 
coupling in comparison to the other MSSM fields that may be non-perturbative.  

Therefore we conclude that it is very difficult to have a viable model
of ``pure'' \gmsb with three degenerate, sequestered MSSM families
within our framework.  Finally recall that realistic type I models do not
predict three degenerate, intersection state families, but instead
distinguish the third family as either $5_1$ or $5_2$ states~\cite{shiutye}.

%%%%%%%%%%%%%%%%%%%%%%%%%%%%%%%%%%%%%%%%%%%%%%%%%%%%%%%%%%%%%%%
%%%%%%%%%%%%%%%%%%% REFERENCES %%%%%%%%%%%%%%%%%%%%%%%%%%%%%%%%
%\newpage


\begin{thebibliography}{99}
%%%%%%%%%%%%%%%%%%%%%%%%%%%%%%%%%%%%%%%%%%%%%%%%%%%%%%%%%%%%%%%
\bibitem{dual}
A.~Font, L.~E.~Ibanez, D.~Lust and F.~Quevedo,
%``Strong-Weak Coupling Duality And Nonperturbative Effects In String 
%Theory,''
Phys.\ Lett.\ B {\bf 249} (1990) 35;
%%CITATION = PHLTA,B249,35;%%
%
M.~J.~Duff and J.~X.~Lu,
%``String/Five-Brane Duality, Loop Expansions And The Cosmological Constant,''
Nucl.\ Phys.\ B {\bf 357} (1991) 534;
%%CITATION = NUPHA,B357,534;%%
%
A.~Sen,
%``Strong - weak coupling duality in four-dimensional string theory,''
Int.\ J.\ Mod.\ Phys.\ A {\bf 9} (1994) 3707
[arXiv:hep-th/9402002].
%%CITATION = HEP-TH 9402002;%%
%%%%%%%%%%%%%%%%%%%%%%%%%%%%%%%%%%%%%%%%%%%%%%%%%%%%%%%%%%%%%%%
\bibitem{mtheory}
C.~M.~Hull and P.~K.~Townsend,
%``Unity of superstring dualities,''
Nucl.\ Phys.\ B {\bf 438} (1995) 109
[arXiv:hep-th/9410167];
%%CITATION = HEP-TH 9410167;%%
%
E.~Witten,
%``String theory dynamics in various dimensions,''
Nucl.\ Phys.\ B {\bf 443} (1995) 85
[arXiv:hep-th/9503124].
%%CITATION = HEP-TH 9503124;%%
%%%%%%%%%%%%%%%%%%%%%%%%%%%%%%%%%%%%%%%%%%%%%%%%%%%%%%%%%%%%%%%
\bibitem{dbranes}
For a review, see:
J.~Polchinski,
%``TASI lectures on D-branes,''
arXiv:hep-th/9611050;
%%CITATION = HEP-TH 9611050;%%
%
C.~V.~Johnson,
%``D-brane primer,''
arXiv:hep-th/0007170.
%%CITATION = HEP-TH 0007170;%%
%%%%%%%%%%%%%%%%%%%%%%%%%%%%%%%%%%%%%%%%%%%%%%%%%%%%%%%%%%%%%%%
\bibitem{largexd}
I.~Antoniadis,
%``A Possible New Dimension At A Few Tev,''
Phys.\ Lett.\ B {\bf 246} (1990) 377;
%%CITATION = PHLTA,B246,377;%%
%
J.~D.~Lykken,
%``Weak Scale Superstrings,''
Phys.\ Rev.\ D {\bf 54} (1996) 3693
[arXiv:hep-th/9603133].
%%CITATION = HEP-TH 9603133;%%
%%%%%%%%%%%%%%%%%%%%%%%%%%%%%%%%%%%%%%%%%%%%%%%%%%%%%%%%%%%%%%%
\bibitem{xduni}
K.~R.~Dienes, E.~Dudas and T.~Gherghetta,
%``Extra spacetime dimensions and unification,''
Phys.\ Lett.\ B {\bf 436} (1998) 55
[arXiv:hep-ph/9803466];
%%CITATION = HEP-PH 9803466;%%
%
%``Grand unification at intermediate mass scales through extra dimensions,''
Nucl.\ Phys.\ B {\bf 537} (1999) 47
[arXiv:hep-ph/9806292].
%%CITATION = HEP-PH 9806292;%%
%%%%%%%%%%%%%%%%%%%%%%%%%%%%%%%%%%%%%%%%%%%%%%%%%%%%%%%%%%%%%%%
\bibitem{xdrev}
For recent reviews, see:
V.~A.~Rubakov,
%``Large and infinite extra dimensions: An introduction,''
Phys.\ Usp.\ {\bf 44} (2001) 871
[Usp.\ Fiz.\ Nauk {\bf 171} (2001) 913]
[arXiv:hep-ph/0104152];
%%CITATION = HEP-PH 0104152;%%
%
Y.~A.~Kubyshin,
%``Models with extra dimensions and their phenomenology,''
arXiv:hep-ph/0111027.
%%CITATION = HEP-PH 0111027;%%
%%%%%%%%%%%%%%%%%%%%%%%%%%%%%%%%%%%%%%%%%%%%%%%%%%%%%%%%%%%%%%%
\bibitem{xdrev2}
M.~Quiros,
%``New ideas in symmetry breaking,''
arXiv:hep-ph/0302189.
%%CITATION = HEP-PH 0302189;%%
%%%%%%%%%%%%%%%%%%%%%%%%%%%%%%%%%%%%%%%%%%%%%%%%%%%%%%%%%%%%%%%
\bibitem{hw}
P.~Horava and E.~Witten,
%``Heterotic and type I string dynamics from eleven dimensions,''
Nucl.\ Phys.\ B {\bf 460} (1996) 506
[arXiv:hep-th/9510209];
%%CITATION = HEP-TH 9510209;%%
%
E.~Witten,
%``Strong Coupling Expansion Of Calabi-Yau Compactification,''
Nucl.\ Phys.\ B {\bf 471} (1996) 135
[arXiv:hep-th/9602070];
%%CITATION = HEP-TH 9602070;%%
%
P.~Horava and E.~Witten,
%``Eleven-Dimensional Supergravity on a Manifold with Boundary,''
Nucl.\ Phys.\ B {\bf 475} (1996) 94
[arXiv:hep-th/9603142].
%%CITATION = HEP-TH 9603142;%%
%%%%%%%%%%%%%%%%%%%%%%%%%%%%%%%%%%%%%%%%%%%%%%%%%%%%%%%%%%%%%%%
\bibitem{amsb} 
L.~Randall and ~R.Sundrum, 
%``Out of this world supersymmetry breaking,''
Nucl.\ Phys.\ B {\bf 557} (1999) 79
[arXiv:hep-th/9810155].
%%CITATION = HEP-TH 9810155;%%
%%%%%%%%%%%%%%%%%%%%%%%%%%%%%%%%%%%%%%%%%%%%%%%%%%%%%%%%%%%%%%%
\bibitem{gmsb}
D.~E.~Kaplan, G.~D.~Kribs and M.~Schmaltz,
%``Supersymmetry breaking through transparent extra dimensions,''
Phys.\ Rev.\ D {\bf 62} (2000) 035010 
[arXiv:hep-ph/9911293];
%%CITATION = HEP-PH 9911293;%%
%
Z.~Chacko, M.~A.~Luty, A.~E.~Nelson and E.~Ponton,
%``Gaugino mediated supersymmetry breaking,''
JHEP {\bf 0001} (2000) 003 
[arXiv:hep-ph/9911323].
%%CITATION = HEP-PH 9911323;%%
%%%%%%%%%%%%%%%%%%%%%%%%%%%%%%%%%%%%%%%%%%%%%%%%%%%%%%%%%%%%%%%
\bibitem{noscale}
J.~R.~Ellis, K.~Enqvist and D.~V.~Nanopoulos,
%``A Very Light Gravitino In A No Scale Model,''
Phys.\ Lett.\ B {\bf 147} (1984) 99;
%%CITATION = PHLTA,B147,99;%%
%
J.~R.~Ellis, C.~Kounnas and D.~V.~Nanopoulos,
%``No Scale Supersymmetric Guts,''
Nucl.\ Phys.\ B {\bf 247} (1984) 373.
%%CITATION = NUPHA,B247,373;%%
%%%%%%%%%%%%%%%%%%%%%%%%%%%%%%%%%%%%%%%%%%%%%%%%%%%%%%%%%%%%%%%
\bibitem{dine}
A.~Anisimov, M.~Dine, M.~Graesser and S.~Thomas,
%``Brane world SUSY breaking,''
Phys.\ Rev.\ D {\bf 65} (2002) 105011 
[arXiv:hep-th/0111235];
%%CITATION = HEP-TH 0111235;%%
%
%``Brane world SUSY breaking from string/M theory,''
JHEP {\bf 0203} (2002) 036 
[arXiv:hep-th/0201256].
%%CITATION = HEP-TH 0201256;%%
%%%%%%%%%%%%%%%%%%%%%%%%%%%%%%%%%%%%%%%%%%%%%%%%%%%%%%%%%%%%%%%
\bibitem{benakli}
K.~Benakli,
%``A note on new sources of gaugino masses,''
Phys.\ Lett.\ B {\bf 475} (2000) 77
[arXiv:hep-ph/9911517].
%%CITATION = HEP-PH 9911517;%%
%%%%%%%%%%%%%%%%%%%%%%%%%%%%%%%%%%%%%%%%%%%%%%%%%%%%%%%%%%%%%%%
\bibitem{tmsb}
S.~F.~King and D.~A.~J.~Rayner,
%``Twisted moduli and supersymmetry breaking,''
JHEP {\bf 0207} (2002) 047
[arXiv:hep-ph/0111333];
%%CITATION = HEP-PH 0111333;%%
%
%``Twisted moduli and supersymmetry breaking,''
arXiv:hep-ph/0211242.
%%CITATION = HEP-PH 0211242;%%
%%%%%%%%%%%%%%%%%%%%%%%%%%%%%%%%%%%%%%%%%%%%%%%%%%%%%%%%%%%%%%%
\bibitem{ibanez94}
A.~Brignole, L.~E.~Ibanez and C.~Munoz,
%``Towards a theory of soft terms for the supersymmetric Standard 
%Model,''
Nucl.\ Phys.\ B {\bf 422} (1994) 125
[Erratum-ibid.\ B {\bf 436} (1995) 747]
[arXiv:hep-ph/9308271];
%%CITATION = HEP-PH 9308271;%%
%
A.~Brignole, L.~E.~Ibanez, C.~Munoz and C.~Scheich,
%``Some issues in soft SUSY breaking terms from dilaton / moduli sectors,''
Z.\ Phys.\ C {\bf 74} (1997) 157
[arXiv:hep-ph/9508258].
%%CITATION = HEP-PH 9508258;%%
%%%%%%%%%%%%%%%%%%%%%%%%%%%%%%%%%%%%%%%%%%%%%%%%%%%%%%%%%%%%%%%
\bibitem{ibanez97}
A.~Brignole, L.~E.~Ibanez and C.~Munoz,
 %``Soft supersymmetry-breaking terms from supergravity and superstring
%models,''
arXiv:hep-ph/9707209.
%%CITATION = HEP-PH 9707209;%
%%%%%%%%%%%%%%%%%%%%%%%%%%%%%%%%%%%%%%%%%%%%%%%%%%%%%%%%%%%%%%%
\bibitem{ibanez98}
L.~E.~Ibanez, C.~Munoz and S.~Rigolin,
%``Aspects of type I string phenomenology,''
Nucl.\ Phys.\ B {\bf 553} (1999) 43
[arXiv:hep-ph/9812397].
%%CITATION = HEP-PH 9812397;%%
%%%%%%%%%%%%%%%%%%%%%%%%%%%%%%%%%%%%%%%%%%%%%%%%%%%%%%%%%%%%%%%
\bibitem{susyrev}
 For example,
S.~P.~Martin,
%''A supersymmetry primer,''
arXiv:hep-ph/9709356;
%%CITATION = HEP-PH 9709356;%%
%
D.~J.~H.~Chung, L.~L.~Everett, G.~L.~Kane, S.~F.~King, J.~Lykken and
 L.~T.~Wang, 
%''The soft supersymmetry-breaking Lagrangian: Theory and
% applications,''
arXiv:hep-ph/0312378.
%%CITATION = HEP-PH 0312378;%%
%%%%%%%%%%%%%%%%%%%%%%%%%%%%%%%%%%%%%%%%%%%%%%%%%%%%%%%%%%%%%%%
\bibitem{shiutye}
G.~Shiu and S.~H.~H.~Tye,
%``TeV scale superstring and extra dimensions,''
Phys.\ Rev.\ D {\bf 58} (1998) 106007
[arXiv:hep-th/9805157].
%%CITATION = HEP-TH 9805157;%%
%%%%%%%%%%%%%%%%%%%%%%%%%%%%%%%%%%%%%%%%%%%%%%%%%%%%%%%%%%%%%%%
\bibitem{bmsb}
S.~F.~King and D.~A.~J.~Rayner,
%``Brane mediated supersymmetry breaking,''
Nucl.\ Phys.\ B {\bf 607} (2001) 77
[arXiv:hep-ph/0012076].
%%CITATION = HEP-PH 0012076;%%
%%%%%%%%%%%%%%%%%%%%%%%%%%%%%%%%%%%%%%%%%%%%%%%%%%%%%%%%%%%%%%%
\bibitem{softsusy}
B.~C.~Allanach,
%``SOFTSUSY: A C++ program for calculating supersymmetric spectra,''
Comput.\ Phys.\ Commun.\  {\bf 143} (2002) 305
[arXiv:hep-ph/0104145].
%%CITATION = HEP-PH 0104145;%%
%See also {\it http://allanach.home.cern.ch/allanach/softsusy.html} \, for
%further details.
%%%%%%%%%%%%%%%%%%%%%%%%%%%%%%%%%%%%%%%%%%%%%%%%%%%%%%%%%%%%%%%
\bibitem{ben}
S.~A.~Abel, B.~C.~Allanach, F.~Quevedo, L.~Ibanez and M.~Klein,
%``Soft SUSY breaking, dilaton domination and intermediate scale string
%models,''
JHEP {\bf 0012} (2000) 026
[arXiv:hep-ph/0005260].
%%CITATION = HEP-PH 0005260;%%
%%%%%%%%%%%%%%%%%%%%%%%%%%%%%%%%%%%%%%%%%%%%%%%%%%%%%%%%%%%%%%%
\bibitem{ben2} 
B.~C.~Allanach, D.~Grellscheid and F.~Quevedo, 
%``Selecting supersymmetric string scenarios from sparticle spectra,'' 
JHEP {\bf 0205} (2002) 048 
[arXiv:hep-ph/0111057];
%%CITATION = HEP-PH 0111057;%%
%
D.~Grellscheid,
%``Required experimental accuracy to select between supersymmetrical  
%models,''
arXiv:hep-ph/0304277.
%%CITATION = HEP-PH 0304277;%%
%%%%%%%%%%%%%%%%%%%%%%%%%%%%%%%%%%%%%%%%%%%%%%%%%%%%%%%%%%%%%%%
\bibitem{gs} 
L.~E.~Ibanez, R.~Rabadan and A.~M.~Uranga, 
%``Anomalous U(1)'s in type I and type IIB D = 4, N = 1 string 
%vacua,'' 
Nucl.\ Phys.\ B {\bf 542} (1999) 112 
[arXiv:hep-th/9808139]. 
%%CITATION = HEP-TH 9808139;%%
%%%%%%%%%%%%%%%%%%%%%%%%%%%%%%%%%%%%%%%%%%%%%%%%%%%%%%%%%%%%%%%
\bibitem{abel}
S.~A.~Abel and G.~Servant,
%``Dilaton stabilization in effective type I string models,''
Nucl.\ Phys.\ B {\bf 597} (2001) 3
[arXiv:hep-th/0009089];
%%CITATION = HEP-TH 0009089;%%
%
%``CP and flavor in effective type I string models,''
Nucl.\ Phys.\ B {\bf 611} (2001) 43
[arXiv:hep-ph/0105262].
%%CITATION = HEP-PH 0105262;%%
%%%%%%%%%%%%%%%%%%%%%%%%%%%%%%%%%%%%%%%%%%%%%%%%%%%%%%%%%%%%%%%
\bibitem{kobayashi}
T.~Higaki and T.~Kobayashi,
%``Twisted moduli stabilization in type I string models,''
Phys.\ Rev.\ D {\bf 68} (2003) 046006
[arXiv:hep-th/0304200];
%%CITATION = HEP-TH 0304200;%%
%
T.~Kobayashi and O.~Seto,
%``Dilaton and moduli fields in D-term inflation,''
Phys.\ Rev.\ D {\bf 69} (2004) 023510
[arXiv:hep-ph/0307332];
%%CITATION = HEP-PH 0307332;%%
%
T.~Higaki, Y.~Kawamura, T.~Kobayashi and H.~Nakano,
%``Anomalous U(1) D-term contribution in type I string models,''
arXiv:hep-ph/0308110.
%%CITATION = HEP-PH 0308110;%%
%%%%%%%%%%%%%%%%%%%%%%%%%%%%%%%%%%%%%%%%%%%%%%%%%%%%%%%%%%%%%%%
\bibitem{altexp}
S.~Hamidi and C.~Vafa,
%``Interactions On Orbifolds,''
Nucl.\ Phys.\ B {\bf 279} (1987) 465;
%%CITATION = NUPHA,B279,465;%%
%
L.~J.~Dixon, D.~Friedan, E.~J.~Martinec and S.~H.~Shenker,
%``The Conformal Field Theory Of Orbifolds,''
Nucl.\ Phys.\ B {\bf 282} (1987) 13;
%%CITATION = NUPHA,B282,13;%%
%
L.~E.~Ibanez,
%``Hierarchy Of Quark - Lepton Masses In Orbifold Superstring
%Compactification,''
Phys.\ Lett.\ B {\bf 181} (1986) 269.
%%CITATION = PHLTA,B181,269;%%
%%%%%%%%%%%%%%%%%%%%%%%%%%%%%%%%%%%%%%%%%%%%%%%%%%%%%%%%%%%%%%%
\bibitem{peddie}
S.~F.~King and I.~N.~R.~Peddie,
%``Canonical normalisation and Yukawa matrices,''
arXiv:hep-ph/0312237.
%%CITATION = HEP-PH 0312237;%%
%%%%%%%%%%%%%%%%%%%%%%%%%%%%%%%%%%%%%%%%%%%%%%%%%%%%%%%%%%%%%%%
\bibitem{twistkp}
E.~Poppitz,
%``On the one loop Fayet-Iliopoulos term in chiral four 
%dimensional type I orbifolds,''
Nucl.\ Phys.\ B {\bf 542} (1999) 31
[arXiv:hep-th/9810010];
%%CITATION = HEP-TH 9810010;%%
%
C.~A.~Scrucca and M.~Serone,
%``Sigma-model symmetry in orientifold models,''
JHEP {\bf 0007} (2000) 025
[arXiv:hep-th/0006201];
%%CITATION = HEP-TH 0006201;%%
%%%%%%%%%%%%%%%%%%%%%%%%%%%%%%%%%%%%%%%%%%%%%%%%%%%%%%%%%%%%%%%
\bibitem{ross}
G.~G.~Ross and O.~Vives,
%``Yukawa structure, flavour and CP violation in supergravity,''
Phys.\ Rev.\ D {\bf 67} (2003) 095013
[arXiv:hep-ph/0211279].
%%CITATION = HEP-PH 0211279;%%
%%%%%%%%%%%%%%%%%%%%%%%%%%%%%%%%%%%%%%%%%%%%%%%%%%%%%%%%%%%%%%%
\bibitem{peddie2}
S.~F.~King and I.~N.~R.~Peddie, 
%``Lepton flavour violation from Yukawa operators, supergravity 
%and the see-saw mechanism,''
Nucl.\ Phys.\ B {\bf 678} (2004) 339
[arXiv:hep-ph/0307091].
%%CITATION = HEP-PH 0307091;%%
%%%%%%%%%%%%%%%%%%%%%%%%%%%%%%%%%%%%%%%%%%%%%%%%%%%%%%%%%%%%%%%%
\bibitem{finetune}
G.~L.~Kane and S.~F.~King,
%``Naturalness implications of LEP results,''
Phys.\ Lett.\ B {\bf 451} (1999) 113
[arXiv:hep-ph/9810374];
%%CITATION = HEP-PH 9810374;%%
%
M.~Bastero-Gil, G.~L.~Kane and S.~F.~King,
%``Fine-tuning constraints on supergravity models,''
Phys.\ Lett.\ B {\bf 474} (2000) 103
[arXiv:hep-ph/9910506].
%%CITATION = HEP-PH 9910506;%%
%%%%%%%%%%%%%%%%%%%%%%%%%%%%%%%%%%%%%%%%%%%%%%%%%%%%%%%%%%%%%%%
\bibitem{fcnc}
F.~Gabbiani, E.~Gabrielli, A.~Masiero and L.~Silvestrini,
%``A complete analysis of FCNC and CP constraints in general 
%SUSY extensions of the standard model,''
Nucl.\ Phys.\ B {\bf 477} (1996) 321
[arXiv:hep-ph/9604387].
%%CITATION = HEP-PH 9604387;%%
%%%%%%%%%%%%%%%%%%%%%%%%%%%%%%%%%%%%%%%%%%%%%%%%%%%%%%%%%%%%%%%
\bibitem{pdg}
K.~Hagiwara {\it et al.} [Particle Data Group Collaboration], 
%``Review Of Particle Physics,'' 
Phys.\ Rev.\ D {\bf 66} (2002) 010001. 
%%CITATION = PHRVA,D66,010001;%%
%%%%%%%%%%%%%%%%%%%%%%%%%%%%%%%%%%%%%%%%%%%%%%%%%%%%%%%%%%%%%%%
\bibitem{triangle}
G.~L.~Fogli, E.~Lisi, A.~Marrone and G.~Scioscia, 
%``Super-Kamiokande atmospheric neutrino data, zenith distributions, 
%and three-flavor oscillations,''
Phys.\ Rev.\ D {\bf 59} (1999) 033001
[arXiv:hep-ph/9808205].
%%CITATION = HEP-PH 9808205;%%
%%%%%%%%%%%%%%%%%%%%%%%%%%%%%%%%%%%%%%%%%%%%%%%%%%%%%%%%%%%%%%%
\bibitem{snowmass}
B.~C.~Allanach {\it et al.},
%``The Snowmass points and slopes: Benchmarks for SUSY searches,''
in {\it Proc. of the APS/DPF/DPB Summer Study on the Future of 
Particle Physics (Snowmass 2001) } ed. N.~Graf,
Eur.\ Phys.\ J.\ C {\bf 25} (2002) 113
[eConf {\bf C010630} (2001) P125]
[arXiv:hep-ph/0202233].
%%CITATION = HEP-PH 0202233;%%
%%%%%%%%%%%%%%%%%%%%%%%%%%%%%%%%%%%%%%%%%%%%%%%%%%%%%%%%%%%%%%%
\bibitem{kane}
G.~L.~Kane, J.~Lykken, S.~Mrenna, B.~D.~Nelson, L.~T.~Wang and T.~T.~Wang,
 %``Theory-motivated benchmark models and superpartners at the Tevatron.
%((V)),''
Phys.\ Rev.\ D {\bf 67} (2003) 045008
[arXiv:hep-ph/0209061].
%%CITATION = HEP-PH 0209061;%%
%%%%%%%%%%%%%%%%%%%%%%%%%%%%%%%%%%%%%%%%%%%%%%%%%%%%%%%%%%%%%%%
\bibitem{leshouches}
P.~Skands {\it et al.}
%``SUSY Les Houches accord: Interfacing SUSY spectrum calculators
%decay packages, and event generators,''
arXiv:hep-ph/0311123.
%%%%%%%%%%%%%%%%%%%%%%%%%%%%%%%%%%%%%%%%%%%%%%%%%%%%%%%%%%%%%%%
\bibitem{barr}
A.~J.~Barr, C.~G.~Lester, M.~A.~Parker, B.~C.~Allanach and P.~Richardson,
%``Discovering anomaly-mediated supersymmetry at the LHC,''
JHEP {\bf 0303} (2003) 045
[arXiv:hep-ph/0208214].
%%CITATION = HEP-PH 0208214;%%
%%%%%%%%%%%%%%%%%%%%%%%%%%%%%%%%%%%%%%%%%%%%%%%%%%%%%%%%%%%%%%%
\bibitem{baer}
H.~Baer, A.~Belyaev, T.~Krupovnikas and A.~Mustafayev,
%``SUSY Normal Scalar Mass Hierarchy Reconciles (g-2), b->s,gamma
%andRelic Density,''
arXiv:hep-ph/0403214.
%%CITATION = HEP-PH 0403214;%%
%%%%%%%%%%%%%%%%%%%%%%%%%%%%%%%%%%%%%%%%%%%%%%%%%%%%%%%%%%%%%%%
\end{thebibliography}
\end{document}